\documentclass[twocolumn]{aastex63}
\usepackage{xspace}
\usepackage{amsmath}

\defcitealias{Poggianti2017}{Paper I}
\defcitealias{Bellhouse2017}{Paper II}
\defcitealias{Fritz2017}{Paper III}
\defcitealias{Gullieuszik2017}{Paper IV}
\defcitealias{Moretti2018}{Paper V}
\defcitealias{Vulcani2018}{Paper VII}
\defcitealias{Vulcani2017c}{Paper VIII}
\defcitealias{Jaffe2018}{Paper IX}
\defcitealias{Vulcani2018b}{Paper XII}
\defcitealias{Vulcani2018_L}{Paper XIV}
\defcitealias{Bellhouse2019}{Paper XV}
\defcitealias{Poggianti2019}{Paper XIII}
\defcitealias{Vulcani2018c}{Paper XIV}
\defcitealias{Vulcani2019b}{Paper XX}
\defcitealias{Poggianti2019b}{Paper XXIII}

\newcommand{\Ha}{H$\alpha$\xspace}

\newcommand{\kms}{$\rm km \, s^{-1}$\xspace}
\newcommand{\kube}{{\sc kubeviz}\xspace}
\newcommand{\sinopsis}{{\sc sinopsis}\xspace}
\newcommand{\ma}{$\rm M_\ast$\xspace}
\newcommand{\ms}{$\rm M_\odot$\xspace}
\newcommand{\Ssfr}{$\rm \Sigma_{SFR}$\xspace}
\newcommand{\Sm}{$\rm \Sigma_\ast$\xspace}

\submitjournal{ApJ}

\shorttitle{The  spatially resolved SFR-Mass relation in stripping galaxies}
\shortauthors{B. Vulcani et al.}

\begin{document}

\title{GASP XXX. The spatially resolved SFR-Mass relation in stripping galaxies in the local universe}

\correspondingauthor{Benedetta Vulcani}
\email{benedetta.vulcani@inaf.it}

\author[0000-0003-0980-1499]{Benedetta Vulcani}
\affiliation{INAF- Osservatorio astronomico di Padova, Vicolo Osservatorio 5, 35122 Padova, Italy}

\author{Bianca M. Poggianti}
\affiliation{INAF- Osservatorio astronomico di Padova, Vicolo Osservatorio 5, 35122 Padova, Italy}

\author{Stephanie Tonnesen}
\affiliation{Center for Computational Astrophysics, Flatiron Institute, 162 5th Ave, New York, NY 10010, USA}

\author{Sean~L. McGee},
\affiliation{University of Birmingham School of Physics and Astronomy, Edgbaston, Birmingham, United Kingdom}

\author{Alessia Moretti}
\affiliation{INAF- Osservatorio astronomico di Padova, Vicolo Osservatorio 5, 35122 Padova, Italy}

\author{Jacopo Fritz}
\affiliation{Instituto de Radioastronom\'ia y Astrof\'isica, UNAM, Campus Morelia, A.P. 3-72, C.P. 58089, Mexico}

\author{Marco Gullieuszik }
\affiliation{INAF- Osservatorio astronomico di Padova, Vicolo Osservatorio 5, 35122 Padova, Italy}

\author{Yara~L. Jaff\'e }
\affiliation{Instituto de F\'isica y Astronom\'ia, Universidad de Valpara\'iso, Avda. Gran Breta\~na 1111 Valpara\'iso, Chile}

\author{Andrea Franchetto}
\affiliation{Dipartimento di Fisica \& Astronomia ``Galileo Galilei'', Universit\`a di Padova, vicolo dell' Osservatorio 3, 35122, Padova, Italy}
\affiliation{INAF- Osservatorio astronomico di Padova, Vicolo Osservatorio 5, 35122 Padova, Italy}

\author{Neven Tomi\v{c}i\'{c}}
\affiliation{INAF- Osservatorio astronomico di Padova, Vicolo Osservatorio 5, 35122 Padova, Italy}

\author{Matilde Mingozzi}
\affiliation{INAF- Osservatorio astronomico di Padova, Vicolo Osservatorio 5, 35122 Padova, Italy}

\author{Daniela Bettoni}
\affiliation{INAF- Osservatorio astronomico di Padova, Vicolo Osservatorio 5, 35122 Padova, Italy}

\author{Anna Wolter}
\affiliation{INAF-Osservatorio Astronomico di Brera, via Brera 28, I-20121 Milano, Italy}

\begin{abstract}
The study of the spatially resolved Star Formation Rate-Mass (\Ssfr-\Sm) relation gives important insights on how galaxies assemble at different spatial scales. Here we present the analysis of the \Ssfr-\Sm of 40 local cluster galaxies undergoing ram pressure stripping drawn from the GAs Stripping Phenomena in galaxies (GASP) sample. Considering their integrated properties, these galaxies 
{ show} a SFR enhancement with respect to undisturbed galaxies of similar stellar mass; we now exploit spatially resolved data to investigate the origin and location of the excess. Even on $~\sim1$kpc scales, stripping galaxies present a systematic enhancement of \Ssfr ($\sim 0.35$ dex at \Sm=$\rm 10^{8}M_\odot \,kpc^{-2}$) at any given \Sm compared to their undisturbed counterparts. The excess is independent on the degree of stripping and of the amount of star formation in the tails and it is visible at all galactocentric distances within the disks, suggesting that the star formation is most likely induced by compression waves from ram pressure. Such excess is larger for less massive galaxies and decreases with increasing mass.
As stripping galaxies are characterised by ionised gas beyond the stellar disk, we also investigate the  properties of 411 star forming clumps found in the galaxy tails. At any given stellar mass density, these clumps are systematically forming  stars at a higher rate than in the disk, but differences are reconciled when we just consider the mass formed in the last few 10$^8$yr ago, suggesting that on these timescales the local mode of star formation is similar in the tails and in the disks. 
\end{abstract}

\keywords{galaxies: clusters: general  --- galaxies: evolution --- galaxies: formation
 --- galaxies: general --- galaxies: star formation}

\section{Introduction}
The existing correlation between a { galaxy's} stellar mass (\ma)  and its ongoing Star Formation Rate (SFR) is one of the most widely studied relations in modern astrophysics \citep[see][for a compilation]{Speagle2014}. Specifically, it relates the stars that have been formed throughout the entire galaxy life to the ongoing SFR, allowing us to investigate the process of star formation and thus  galaxy evolution as a whole. Overall, among star-forming galaxies,  higher stellar mass systems undergo more intense star formation activity than lower  mass  systems \citep[e.g.][]{Noeske2007b}.
The existence of such relation { and especially its low dispersion (0.2-0.3 dex at all redshifts)} points to a scenario where galaxies form through  secular processes rather than stochastic merger-driven star-forming episodes. Their evolution throughout cosmological time and across environments must therefore be regulated by the same universal laws \citep[e.g.,][]{Noeske2007a, Noeske2007b, Bouche2010, Daddi2010, Genzel2010, Tacconi2010,  Dave2012,  Dayal2013, Dekel2013, Lilly2013, Feldmann2015, Tacchella2016}.

{ Only galaxies with SFRs well above the main sequence \citep{Rodighiero2011, Rodighiero2015,  Silverman2015} contrast this picture as they can be interpreted as evidence of starbursts triggered by mergers or external inflows; however, the recent observational evidence on the young age of these systems \citep[e.g.,][]{daCunha2015, Ma2015} points toward an alternative interpretation in line with the in situ scenario.}

The MS  has been studied for the first time by \cite{Brinchmann2004} for local galaxies and later confirmed for high-redshift galaxies by several works \citep[e.g.,][]{Salim2007, Noeske2007b, Elbaz2007, Daddi2007, Speagle2014, Schreiber2015, Kurczynski2016, Santini2017, Tacchella2016, Pearson2018, Popesso2019, Morselli2019}. These studies are based on integrated quantities, therefore consider galaxies as a whole, not always distinguishing among the morphological components or excluding regions hosting Active Galactic Nuclei (AGN), if any. 
In addition, many observations cover only partially the optical extent of the galaxies and are thus subject to aperture effects.

To overcome these issues, in  recent years, efforts have been devoted to analyze the { SFR-\ma} relation  on smaller scales, using spatially resolved data, therefore characterizing the so called ``local'' relation, in contrast
to the ``global'' one based on integrated properties. 

{ The comparison between the local and global relation can shed light on the interplay between different galaxy scales, i.e. on the physical processes connecting local parameters of star formation and feedback to the global star formation in galaxies \citep[e.g.,][]{Semenov2018}. It can also help to determine the minimum scale at which the mechanism that drives the star formation activity with respect to the stellar mass could be universal.}

While some studies attempted to characterise the small scales using photometric data \citep{Abdurrouf2017, Abdurrouf2018, Morselli2018, Hemmati2020}, the great step forward for this kind of analysis has been possible thanks to the advent of large integral field spectroscopic (IFS) surveys \citep[e.g.][]{Sanchez2012, Bundy2015,Bryant2015}.
All  studies report the existence of a correlation even at smaller scales (down to the sizes of molecular clouds), thus implying that the star formation process is regulated by physical processes that act on sub-galactic scales \citep{RosalesOrtega2012, Sanchez2013, CanoDiaz2016, Hsieh2017, Lin2017, Pan2018, Liu2018, Medling2018, Hall2018, Erroz2019, CanoDiaz2019, Vulcani2019b, Bluck2020, Enia2020, Morselli2020}. Nonetheless,  the slope, intercept and scatter of the relation vary significantly among different works. These discrepancies are most likely due to the different  sample selection, star formation indicator, dust correction, and fitting procedure adopted by the various authors. 

Moreover, some authors highlight  that the spatially resolved relation varies dramatically from galaxy to galaxy \citep[e.g.,][]{Hall2018, Vulcani2019b} and that some specific galaxy populations can deviate from the general relation \citep[e.g.][]{CanoDiaz2019, Medling2018}. 
Investigating which specific populations do not follow the general trends, both on local and global scales, can give useful insight on their evolution.

\cite{CanoDiaz2019, Medling2018} have found that galaxies of different { global} morphology  occupy distinct loci on the { spatially resolved} MS: the earlier the morphological type, the lower is on average the  { spatially resolved SFR (\Ssfr)}, even for galaxies of similar { spatially resolved \ms (\Sm)}. Similarly, also on global scales the SFR-mass relation depends on morphology, with late-type galaxies having systematically higher SFR values than early types \citep[e.g.,][]{Calvi2018MorphologyUniverse}. 

\cite{Ellison2020} have shown that variations in Star Formation Efficiency \citep[SFE = \Ssfr/$\Sigma_{\rm H2}$, e.g.,][]{Genzel2015} are responsible for variations in \Ssfr on kpc-scales, therefore galaxies that highly deviate  from the fit of the relation have very high SFE. Similar results have been obtained also using global  quantities: \cite{Genzel2015, Silverman2015, Silverman2018,  Tacconi2018} showed that there is a correlation between the position of a galaxy relative to the MS ($\Delta$SFR) and its total SFE. 
{ 
\cite{Saintonge2012, Saintonge2016} have highlighted also a dependence on the global 
 cold gas reservoirs, with in addition systematic variations in the molecular-to-atomic ratio. However, \cite{Ellison2020} have shown that on local scales the dependence on gas fraction 
is only secondary to the SFE and  weaker.}

Also galaxies in the densest environments have been shown to deviate from the general field population, on global scales. Both at $z=0$ and up to $z\sim1$, cluster galaxies can be as star forming as field galaxies, but a population of galaxies with a suppressed SFR at any given mass has been detected \citep{Vulcani2010ComparingEnvironments, Paccagnella2016, Guglielmo2019, Old2020}. On local scales, only \cite{Vulcani2019b} have compared the local MS of galaxies in clusters and field, but the way their sample was assembled (i.e. morphologically undisturbed star-forming galaxies on the global MS) prevented them from investigating eventual differences in the large population of morphologically disturbed galaxies in clusters.

These results suggest that the local star formation (at scales $>\sim$1 kpc$^2$) is established by some universal process, but it is modulated partially by global properties, such as the  morphology of the galaxy or the gas fraction.

The local MS has been shown also  to drive the global one  \citep[see also][]{Hsieh2017, CanoDiaz2016}, most likely through the existence of the size-mass relation: on local scales the mean \Ssfr and \Sm values for all galaxies are quite similar, regardless of the galaxy size, while on global scales more extended galaxies are also more massive and more star-forming \citep{Vulcani2019b}.

In this context, another population worth investigating are the cluster galaxies that are currently losing their gas via ram-pressure stripping (RPS) due to their motion through the intracluster medium \citep[ICM;][]{Gunn1972},
before being fully quenched \citep{Vulcani2020}.
The most spectacular examples of galaxies losing gas are the so-called jellyfish galaxies. They are at the peak of the stripping and show tails with ionized gas and bright blue knots downstream of the disks, indicating substantial SF in their tails, and asymmetric disks of young stars \citep[e.g.,][]{Cortese2007, Smith2010, Fumagalli2014, Fossati2016, Consolandi2017, Poggianti2017, Moretti2018, Gullieuszik2017, Bellhouse2017, Boselli2018}.

Observationally, { it has been shown that} RPS generally enhances the star formation before quenching it (\citealt{Crowl2006,  Merluzzi2013, Kenney2014}, but see  \citealt{Crowl2008} for a different interpretation). \cite{Vulcani2018_L} showed that stripping galaxies lay above the SFR-\ma relation of undisturbed galaxies, indicating that star formation is boosted in the disks during stripping.  Additional star formation takes place in the tails (see also \citealt{Fumagalli2014, Poggianti2017, Roman2019, Cramer2018, Boselli2018}, but \citealt{Boselli2016} for different results). 
This observed enhancement is linked to the higher molecular gas reservoir these galaxies have. Moretti et al. (submitted) have indeed shown that galaxies at peak stripping are very efficient in converting HI into H$_2$.

Simulations overall support the observational results (\citealt{Kronberger2008, Kapferer2009}, but \citealt{TonnesenBryan2012} did not find a significant star formation enhancement), even though they are not always concordant on the portion of the galaxy which shows the enhancement.  \cite{Kronberger2008} found that even though new stars are mainly formed in the central parts of the disk, a significant fraction forms also  in the wake of the galaxy, while \cite{Kapferer2009} found a shift in the star formation from the disk to the wake, with a net SFR suppression in the disk. \cite{Roediger2014} showed that star formation enhancements take place only in regions of sufficiently low initial interstellar medium pressure, which will be stripped soon afterward. \cite{Troncoso2020} divided the galaxy with a plane perpendicular to the galaxy velocity direction and found an enhancement  in the half of the galaxy approaching the cluster center. \cite{Bekki2014, Steinhauser2016} found that the enhancement depends strongly on the satellite mass, orbit, and inclination angle. 

In this context, a great step forward on the characterisation and interpretation of stripping galaxies has been possible thanks to the  GAs Stripping Phenomena in galaxies with MUSE (GASP\footnote{\url{http://web.oapd.inaf.it/gasp/index.html}}) project, an ESO Large Programme granted 120hr of observing time with the integral field spectrograph MUSE that was completed in 2018. GASP allows us to study galaxies in the local universe in various stages of RPS in clusters \citep{Jaffe2018} and provides us with the unique possibility of looking for trends and performing comparisons in a homogeneous sample, reducing possible biases. A complete description of the survey can be found in  \citet{Poggianti2017}.

In this paper, we make use of the GASP sample to investigate for the first time the spatially resolved SFR-\ma relation of { galaxies currently being stripped by ram pressure - called from now on ``stripping galaxies'' for brevity}, with the aim of understanding the origin of the global enhancement observed in \citet{Vulcani2018_L}. 
The first part of the paper will focus only on the galaxy disks, excluding the contribution of the galaxy tails. We will therefore compare stripping galaxies to the control sample studied in \citet{Vulcani2019b}, to investigate what drives the observed enhancement and localise where such enhancement is. 
In the second part of the paper we will instead only focus on the tails of the stripping galaxies and characterise the star forming  properties of the clumps detected in the galaxy wakes, complementing the characterisation of the clumps presented in \citet{Poggianti2019}. 

The paper is divided in the following sections. 
Sections \ref{sec:data}  presents the data sample and  analysis, and describes the identification and characterisation of the clumps (Section \ref{sec:clumps}). Section  \ref{sec:results} includes the results: Section \ref{sec:results_disk} focuses on the comparison between the stripping and control sample disk \Ssfr-\Sm relations, Section \ref{sec:results_clumps} includes the study of the \Ssfr-\Sm relations for the clumps.
In Section \ref{sec:disc} we  discuss the results and conclude. 

 We adopt a \cite{Chabrier2003} initial mass function (IMF) in the mass range 0.1-100 M$_{\odot}$. The cosmological constants assumed are $\Omega_m=0.3$, $\Omega_{\Lambda}=0.7$ and H$_0=70$ km s$^{-1}$ Mpc$^{-1}$.

\begin{figure*}
\centering
\includegraphics[scale=0.35]{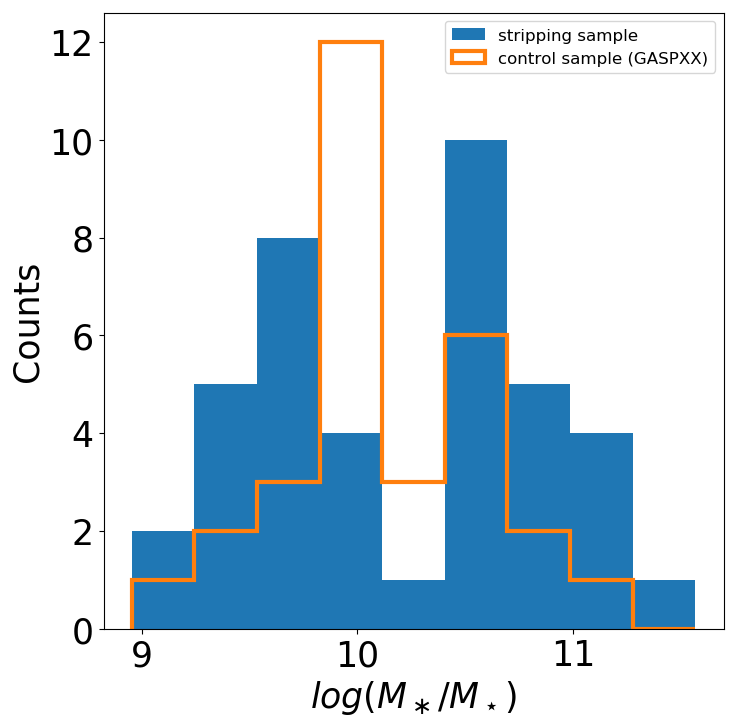}
\includegraphics[scale=0.35]{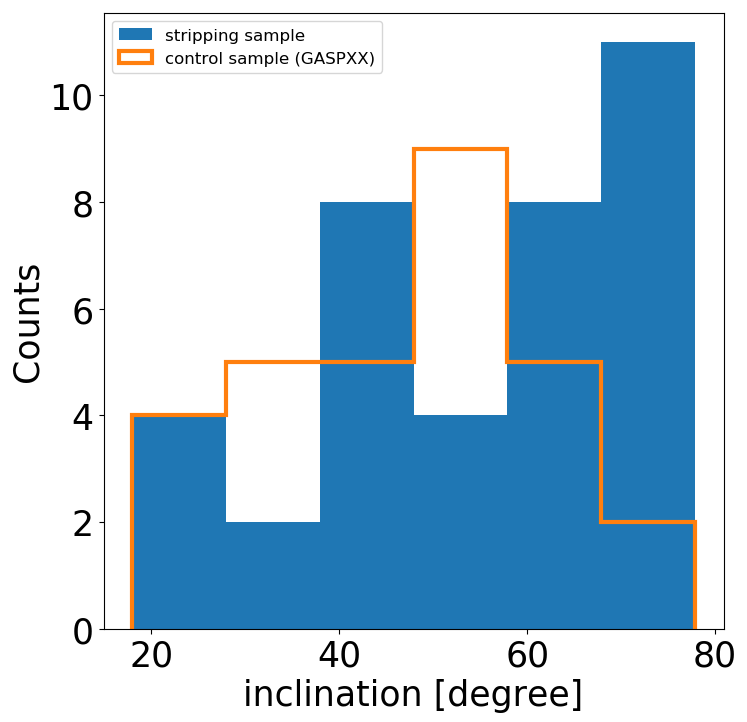}
\caption{Total stellar mass (left) and inclination (right) distribution of the stripping sample (blue) compared to the control sample from \citet{Vulcani2019b} (orange).\label{fig:mass} }
\end{figure*}

\section{Data sample  and  data analysis} \label{sec:data}

 \subsection{Data sample}
 All the observations used in this paper have been obtained in the context of the GASP project. 
The  survey targeted 114 galaxies at redshift $0.04<z<0.1$, spanning a wide range of galaxy stellar masses ($10^9<$\ma/\ms $<10^{11.5}$) and  located in
different environments (galaxy clusters, groups, filaments and
isolated). GASP includes both galaxies selected as stripping candidates and undisturbed galaxies. 

The sample of galaxies analysed in this paper is drawn from the GASP cluster sample and has been presented in \citet{Vulcani2018_L}. Briefly, it includes galaxies with signs of mild, moderate, and extreme stripping, as well as truncated disks. Uncertain cases, as well interacting galaxies { identified on the basis of stellar tails and/or companions in the same field of view}  were disregarded. We further exclude from this sample JO149 and JO95 since for these galaxies effective radii could not be determined (see below). The total stripping sample includes 40 galaxies. We refer to Table 1 of \citet{Vulcani2018_L} for the list of the objects, along with redshifts, coordinates, integrated stellar masses and star formation rates. 

When needed,  we will also use the control sample of galaxies presented in  \citet{Vulcani2018_L} { and already exploited in \citet{Vulcani2019b}}. This sample includes cluster+field galaxies that are undisturbed and do not show any clear  sign of environmental effects (ram pressure stripping, tidal interaction, mergers, gas accretion, or other interactions) on their spatially resolved star formation distribution.  { Similarly to what done in \citet{Vulcani2019b}, we exclude from the sample JO93 and P19482 that, after a careful inspection of their \Ha maps, turned out to be in an initial phase of stripping. The final sample includes } 30  galaxies, 16 of which are cluster members and 14 field galaxies. Table 2 of \citet{Vulcani2018_L} presents the galaxies included in the control sample.  Note that in \citet{Vulcani2019b} we did not find any difference between undisturbed galaxies in clusters and in the field. The result was somehow expected, as those cluster members most likely just entered their  cluster from the field and have had no time yet to feel cluster specific processes.

In \citet{Vulcani2019b} we already compared our results to literature results \citep[e.g.][]{CanoDiaz2016, Hsieh2017}, highlighting how different observational strategies, along with sample selection, analyzing method, fitting recipe and spatial resolution,  play an important role in the determination of the parameters that better describe the relations.  In what follows we will therefore only use our own control sample, which is treated in the exactly same way as our primary sample. 

The left panel of Figure \ref{fig:mass} shows the total stellar mass distribution of the galaxies entering the sample, compared to that of the control sample.\footnote{Colour images  of all GASP galaxies along with \Ha images can be consulted on a webpage at
\url{http://web.oapd.inaf.it/gasp/gasp_atlas}.}

\subsection{Data analysis}\label{sec:analysis}

A complete description of the survey strategy, observations, data reduction and analysis procedure is presented in \citet{Poggianti2017}. 

Briefly, data were reduced with the most recent available version of the MUSE pipeline\footnote{\url{http://www.eso.org/sci/software/pipelines/muse}} and datacubes were  averaged filtered  in the spatial direction with a 5$\times$5 pixel kernel,  corresponding to our worst seeing conditions of 1$^{\prime\prime}$  = 0.7-1.3 kpc at the redshifts of the GASP  galaxies. All the forthcoming results are therefore valid on a scale of $\sim1$ kpc.

We corrected the reduced datacube for extinction due to our Galaxy and subtracted the stellar-only component of each spectrum
derived with our spectrophotometric code \sinopsis \citep{Fritz2017}.  \sinopsis also provides { stellar masses} for each MUSE spaxel.

Emission line fluxes and errors, along with the underlying continuum,  were derived  using the IDL software \kube \citep{Fossati2016}.  We consider as reliable only spaxels with S/N(\Ha)$>$5.
\Ha luminosities corrected both for stellar absorption and for dust
extinction were used to compute SFRs, adopting the \cite{Kennicutt1998a}'s relation: $\rm SFR (M_{\odot}
\, yr^{-1}) = 4.6 \times 10^{-42} L_{\rm H\alpha} (erg \, s^{-1})$. 
The extinction was estimated from the Balmer decrement
assuming a value $\rm H\alpha/H\beta = 2.86$ and the \cite{Cardelli1989} extinction law. 
The MUSE data reach a surface brightness detection limit of $\rm V\sim 27 \, mag \, arcsec^{-2}$ and $\rm  H\alpha \sim 10^{-17.6} \, erg \, s^{-1}\,  cm^{-2} \,  arcsec^{-2}$ at the 3$\sigma$ confidence level \citep{Poggianti2017}, which translates into a \Ssfr limit of $\rm \sim 7 \times 10^{-5} \, M_\odot \, yr^{-1} \, kpc^{-2}$.

We employed the standard diagnostic diagram
[OIII]5007/$\rm H\beta$ vs [OI]6300/$\rm H\alpha$ to separate the regions powered by star formation from regions powered by AGN or LINER emission.\footnote{Among the various line-ratio diagrams, the one based on the [OI] is the most sensitive to physical processes different from Star Formation (e.g. thermal conduction from the surrounding hot ICM, turbulence and shocks) and can therefore be considered as a conservative lower limit of the real star formation budget \citep{Poggianti2019}. In the appendix of \citet{Vulcani2019b} we have shown that results for the control sample are qualitatively independent on the choice of the diagnostic diagram.} Only spaxels with a S/N$>$3 in all emission lines involved are considered. We adopted the division lines by \citet{Kauffmann2003}. For the majority of the galaxies most of the \Ha is powered by photoionization (plots not shown, see e.g. Fig. 2 in \citet{Poggianti2019}), even though 11 galaxies  host an AGN in their center (see M. Radovich et al. in prep.). 
To compute SFRs, we considered only the spaxels whose ionised flux is powered by star formation. 

For each spaxel in each galaxy we also computed the galactocentric radius fixing the centre of the galaxy
to the peak of the stellar mass map. The radius is then expressed in units of  $r_e$, which is computed on I-band images by measuring the radius of an ellipse including half of the total light of the galaxy \citep{Franchetto2020}.  We remind the reader that our observations cover the entire optical extension of the galaxy, up to several effective radii, so our data are not affected by aperture loss.  
All quantities were corrected for the effect of inclination. The inclination distribution of the sample is shown in the right panel of  Fig.\ref{fig:mass}. Galaxies with $i > 70^{\circ}$ will not be excluded from the analysis, even though their results must be taken with caution, so they will be highlighted in the following plots. 

In the following analysis, we will consider separately spaxels within and outside galaxy disks. We use the definition of galaxy boundaries developed by \citet{Gullieuszik2020}. { Briefly, for each galaxy, the galaxy boundaries were estimated by inspecting the stellar isophote corresponding to a surface brightness 1$\sigma$ above the average sky background level. The stellar isophotes were derived by using the continuum map obtained by the KUBEVIZ model of the \Ha+[N II] lineset.
As for stripping galaxies the isophote does not have elliptical symmetry, mainly because of the emission from stars born in the stripped tail, \citet{Gullieuszik2020} fit an ellipse to the undisturbed side of the isophote and used the same ellipse to replace the isophote on the disturbed side. The resulting contour defines a mask that we used to discriminate the galaxy main body and the ram-pressure stripped tail.}
Everything inside of the isophote represents the galaxy disk, the rest constitutes the galaxy tail.

By definition, control sample galaxies have negligible \Ha flux (therefore SFR) in the tails. Therefore, comparisons between the  stripping and the control sample will be performed using only the spaxels belonging to the galaxy disks. 

\subsubsection{Identification and Characterization of \Ha clumps}\label{sec:clumps}
For the stripping sample, we will also investigate 
the properties of  \Ha clumps detected outside the galaxy disks. The clumps have \Ha surface brightness typically between $\rm 10^{-16.5}-10^{-15} \,  erg \, s^{-1} \,  cm^{-2} \,  arcsec^{-2}$. \citet{Poggianti2017} describes in detail how these clumps are identified. Briefly, these are defined by searching 
the local minima of the laplace + median filtered \Ha MUSE image.\footnote{{ Note that the laplacian filtering measures the second  spatial derivative of an image and is commonly defined using a negative peak, that is why we are looking for minima.}} The boundaries of these clumps (i.e. their radius, having assumed circular symmetry) are estimated considering outgoing shells until the average counts reach a threshold value that defines the underlying diffuse emission.

SFRs of the clumps have been computed in the same way as for the single spaxels, only for the clumps whose main ionisation mechanism is photoionisation by young stars, always according to the [OIII]5007/$\rm H\beta$ vs [OI]6300/$\rm H\alpha$ diagrams. \citet{Poggianti2019} showed the BPT diagrams for the clumps in the tails for 16/40 galaxies in the sample. 

Following \citet{Poggianti2019}, stellar mass estimates of clumps in the tails have been obtained running \sinopsis with an upper limit to the age of the stellar populations ($2\times 10^8$ yr). This choice avoids having very low levels of unrealistically old stars in the tails, whose light contribution is insignificant, but whose integrated stellar mass can result in overestimating the stellar mass. In this way, we are more likely to get a fair value of the total stellar mass. Note that \citet{Poggianti2019}  tested that these stellar mass values do not change significantly varying the upper age limit between a few $10^7$ and $10^9$ yr, therefore the measurement is  stable. 

\Ssfr and \Sm are obtained by dividing the  SFR and M$_\ast$  obtained frm the integrated spectrum of the clumps
by the area of the clumps, obtained assuming circular symmetry. 
Note that as both values are divided by the same amount, even in the cases our sizes could be overestimated due to the seeing which is always about 1$^{\prime\prime}$ (see \citet{Poggianti2019}) the correlation in maintained.
In addition, the correlation is valid as long as the tracers of star formation and the stellar mass have the same spatial distribution. 

Our final sample includes 411 \Ha clumps. These ones have been selected for being found outside the stellar disk, are powered by SFR and have S/N$>$3 in all the lines involved in the BPT. 

\begin{table}
\caption{Least-squares regression parameters for the different samples. Note that JO171, J1079 and JW39 are not listed because the fit is not meaningful. The Figure where the relation is used is also listed. Details on the fitting method are given in the text.    \label{tab:param}}
\centering
\begin{tabular}{lrrr}
  \multicolumn{1}{c}{sample} &
  \multicolumn{1}{c}{intercept} &
  \multicolumn{1}{c}{slope} &
  \multicolumn{1}{c}{Fig.} \\
\hline
  stripping (S) & -14.79$^{+0.04}_{-0.05}$ & 1.641$^{+0.006}_{-0.006}$  &  \ref{fig:SFR_Mass_all}, \ref{fig:sfr_mass_dist_mass}, \ref{fig:blobs_sfrd} \\
  control  & -15.28$^{+0.06}_{-0.06}$ & 1.659$^{+0.008}_{-0.008}$ & \ref{fig:SFR_Mass_all}, \ref{fig:SFR_Mass_SFR}, \ref{fig:SFR_Mass_Jstage}, \ref{fig:sfr_mass_dist_mass} \\
  S SFR$_{out}<$0.02 &-13.72$^{+0.08}_{-0.07}$ &1.502$^{+0.009}_{-0.01}$ &\ref{fig:SFR_Mass_SFR}\\
  S SFR$_{out}>$0.02 &-15.10$^{+0.06}_{-0.06}$ &1.681$^{+0.007}_{-0.007}$ &\ref{fig:SFR_Mass_SFR}\\
  S Jstage=0.5 &-15.2$^{+0.1}_{-0.1}$ &1.71$^{+0.02}_{-0.02}$ &\ref{fig:SFR_Mass_Jstage}\\
  S Jstage=1 &-14.00$^{+0.07}_{-0.07}$ &1.544$^{+0.01}_{-0.009}$ &\ref{fig:SFR_Mass_Jstage}\\
  S Jstage=2 &-15.93$^{+0.07}_{-0.07}$ &1.780$^{+0.009}_{-0.01}$ &\ref{fig:SFR_Mass_Jstage}\\
  S Jstage=3 &-18.0$^{+0.2}_{-0.3}$ &1.97$^{+0.03}_{-0.03}$&\ref{fig:SFR_Mass_Jstage}\\
    JO10  & -17.9$^{+ 0.2}_{- 0.2}$   &1.89$^{+0.02}_{-0.02}$ & \ref{fig:sfr_mass_dist_mass}\\
  JO112 &  -19.2$^{+0.4}_{-0.4}$ &  2.28$^{+ 0.05}_{-0.05}$ &  \ref{fig:sfr_mass_dist_mass}\\
  JO113 &  -13.4$^{+0.3}_{-0.3}$   & 1.54$^{+ 0.05}_{- 0.04}$ & \ref{fig:sfr_mass_dist_mass}\\
  JO135 &  -12.9$^{+ 0.2}_{- 0.2}$  &  1.34$^{+0.03}_{-0.02}$ &  \ref{fig:sfr_mass_dist_mass}\\
  JO138 &  -14.1$^{+0.4}_{-0.4}$ &1.56$^{+ 0.06}_{-0.06}$ &  \ref{fig:sfr_mass_dist_mass}\\
  JO13  &  -18.6$^{+0.3}_{- 0.3}$   &  2.20$^{+0.04}_{-0.04}$ &  \ref{fig:sfr_mass_dist_mass}\\
  JO141 & -13.4$^{+ 0.1}_{-0.2}$ & 1.40$^{+ 0.02}_{- 0.02}$ &  \ref{fig:sfr_mass_dist_mass}\\
  JO144 &  -15.0$^{+0.2}_{-0.2}$   &  1.62$^{+0.03}_{-0.03}$ &  \ref{fig:sfr_mass_dist_mass}\\
  JO147 &  -12.2$^{+0.1}_{-0.1}$   &  1.23$^{+ 0.01}_{-0.01}$ &  \ref{fig:sfr_mass_dist_mass}\\
  JO159 &  -16.5$^{+0.5}_{-0.6}$   &  1.87$^{+0.07}_{-0.07}$ &  \ref{fig:sfr_mass_dist_mass}\\
  JO160 &  -16.4$^{+0.2}_{-0.3}$   &  1.87$^{+0.03}_{-0.03}$ &  \ref{fig:sfr_mass_dist_mass}\\
  JO162 &  -11.1$^{+ 0.2}_{-0.3}$   &  1.21$^{+0.04}_{-0.04}$ &  \ref{fig:sfr_mass_dist_mass}\\
  JO175 & -11.78$^{+0.089}_{-0.09}$   &  1.21$^{+0.01}_{-0.01}$ &  \ref{fig:sfr_mass_dist_mass}\\
  JO181 &  -21$^{+1}_{-2}$   &  2.7$^{+0.3}_{-0.2}$ &  \ref{fig:sfr_mass_dist_mass}\\
  JO194 &  -19.6$^{+0.3}_{-0.3}$   &  2.21$^{+0.04}_{-0.04}$ &  \ref{fig:sfr_mass_dist_mass}\\
  JO197 &  -16.4$^{+0.2}_{-0.2}$   &  1.82$^{+0.03}_{-0.03}$ &  \ref{fig:sfr_mass_dist_mass}\\
  JO200 &  -11.9$^{+0.1}_{-0.1}$   &  1.21$^{+0.01}_{-0.01}$ &  \ref{fig:sfr_mass_dist_mass}\\
  JO201 &  -13.8$^{+0.2}_{-0.2}$   &  1.54$^{+0.02}_{-0.02}$ &  \ref{fig:sfr_mass_dist_mass}\\
  JO204 &  -11.2$^{+0.1}_{-0.1}$   &  1.15$^{+0.02}_{-0.02}$ &  \ref{fig:sfr_mass_dist_mass}\\
  JO206 &  -21.1$^{+0.6}_{-0.6}$   &  2.48$^{+0.08}_{-0.07}$ &  \ref{fig:sfr_mass_dist_mass}\\
  JO23  &  -26.1$^{+0.8}_{-0.9}$   &  3.1$^{+0.1}_{-0.1}$ &  \ref{fig:sfr_mass_dist_mass}\\
  JO27  &  -12.3$^{+0.2}_{-0.2}$   &  1.37$^{+0.04}_{-0.03}$ &  \ref{fig:sfr_mass_dist_mass}\\
  JO28  &  -19$^{+1}_{-1}$   &  2.2$^{+0.2}_{-0.1}$ &  \ref{fig:sfr_mass_dist_mass}\\
  JO36  &  -28$^{+1}_{-1}$   &  3.3$^{+0.1}_{-0.1}$ &  \ref{fig:sfr_mass_dist_mass}\\
  JO47  &  -21.3$^{+0.8}_{-0.8}$   &  2.6$^{+0.1}_{-0.1}$ &  \ref{fig:sfr_mass_dist_mass}\\
  JO49  &  -13.5$^{+0.2}_{-0.2}$   &  1.41$^{+0.03}_{-0.03}$ &  \ref{fig:sfr_mass_dist_mass}\\
  JO60  &  -11.2$^{+0.1}_{-0.1}$   &  1.23$^{+0.02}_{-0.02}$ &  \ref{fig:sfr_mass_dist_mass}\\
  JO69  &  -16.9$^{+0.4}_{-0.4}$   &  2.00$^{+0.06}_{-0.05}$ &  \ref{fig:sfr_mass_dist_mass}\\
  JO70  &  -10.11$^{+0.09}_{-0.09}$ &  1.04$^{+0.01}_{-0.01}$ &  \ref{fig:sfr_mass_dist_mass}\\
  JO85  &  -13.6$^{+0.1}_{-0.1}$   &  1.51$^{+0.02}_{-0.02}$ &  \ref{fig:sfr_mass_dist_mass}\\
  JO93  &  -14.2$^{+0.2}_{-0.2}$   &  1.54$^{+0.02}_{-0.02}$ &  \ref{fig:sfr_mass_dist_mass}\\
  JW100 &  -11.9$^{+0.2}_{-0.2}$   &  1.19$^{+0.03}_{-0.03}$ &  \ref{fig:sfr_mass_dist_mass}\\
  JW108 &  -25.5$^{+0.5}_{-0.5}$   &  2.88$^{+0.06}_{-0.06}$ &  \ref{fig:sfr_mass_dist_mass}\\
  JW10  &  -24$^{+1}_{-2}$   &  3.0$^{+0.2}_{-0.2}$ &  \ref{fig:sfr_mass_dist_mass}\\
  JW115 &  -13.7$^{+0.8}_{-0.9}$   &  1.545$^{+0.1}_{-0.1}$ &  \ref{fig:sfr_mass_dist_mass}\\
  JW29  &  -24.2$^{+0.9}_{-1}$   &  3.1$^{+0.1}_{-0.1}$ &  \ref{fig:sfr_mass_dist_mass}\\
  JW56  &  -11.9$^{+0.6}_{-0.7}$   &  1.30$^{+0.1}_{-0.09}$ &  \ref{fig:sfr_mass_dist_mass}\\
  clumps out &  -13.0$^{+0.5}_{-0.5}$  &  1.63$^{+0.08}_{-0.07}$ &  \ref{fig:blobs_sfr} \\
  global stripping  &  -10$^{+2}_{-5}$   &  0.9$^{+0.5}_{-0.2}$ &  \ref{fig:blobs_sfr} \\
\hline\end{tabular}
\end{table}

\begin{figure}
\centering
\includegraphics[scale=0.33]{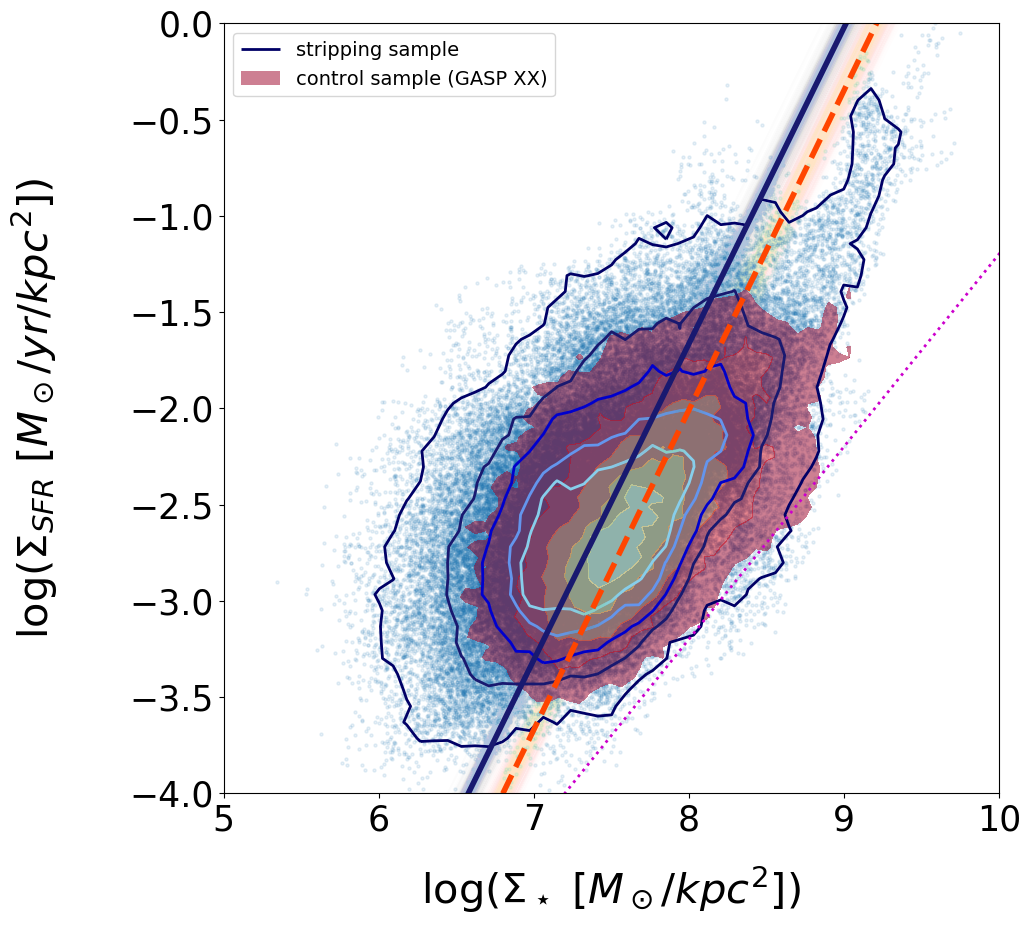}
\caption{Spatially resolved SFR-\ma (\Ssfr - \Sm) relation for all spaxels in the galaxy disks of all stripping  galaxies (blue). Superimposed in a blue scale are contours representing the 15th, 35th, 65th, 85th and 98th percentiles. Superimposed in red{ - to yellow} scale as shaded areas are contours representing the same percentiles for the control sample discussed in \citet{Vulcani2019b}. Thick blue and dashed red  lines show the fit to the relation, for the stripping and control sample, respectively. Transparent lines show samples from the posterior, indicating the scatter in the fit. The magenta dotted line represents the effective threshold in spatially resolved specific SFR entailed by the adopted cuts in S/N and corresponds to $\rm 10^{-11.2}\, yr^{-1} \, kpc^{-2}$. Galaxies in the stripping sample have a systematically higher \Ssfr at any given \Sm than galaxies in the control sample.  \label{fig:SFR_Mass_all} }
\end{figure}

\section{Results} \label{sec:results}
\subsection{Disk \Ssfr - \Sm relation for stripping and control sample galaxies}  \label{sec:results_disk}

\subsubsection{The disk \Ssfr- \Sm relation of all galaxies}

\begin{figure*}
\centering
\includegraphics[scale=0.45]{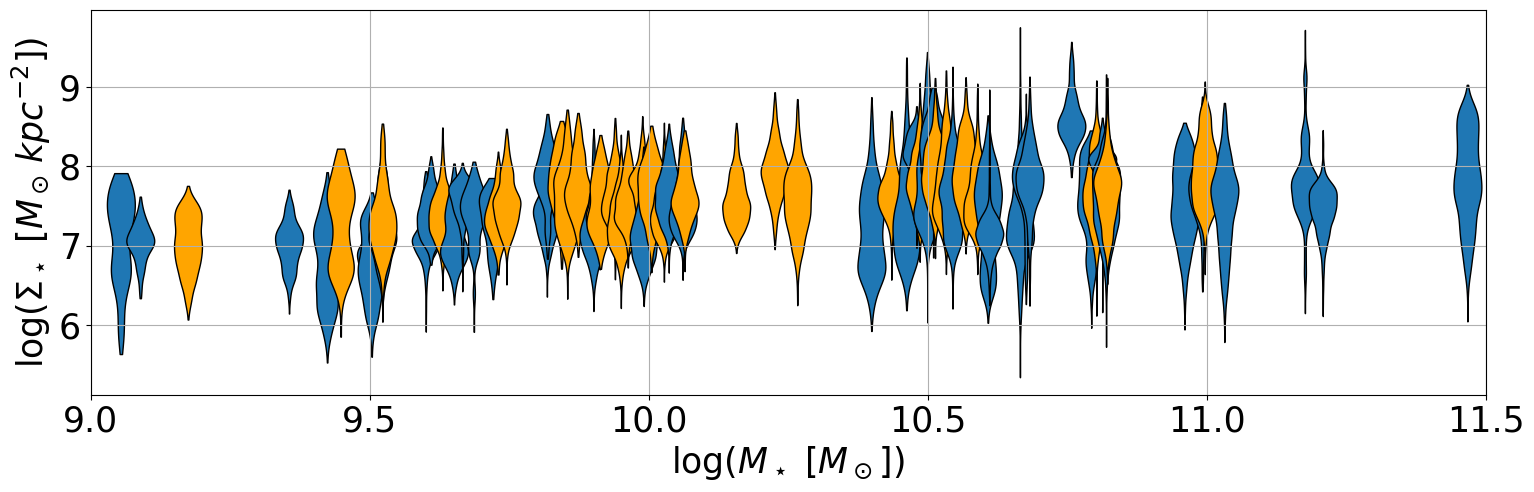}
\caption{Violin plots of the \Sm distribution of the the galaxies in the stripping (blue) and control (orange) sample, as a function of the galaxy stellar mass.  \Sm range depends on $M_\ast$. Galaxies with $M_\ast <10^{9.75} M_\odot$ have \Sm always lower than 10$^8 M_\odot kpc^{-2}$, while in massive galaxies \Sm can reach and even exceed 10$^9 M_\odot kpc^{-2}$. 
\label{fig:violin_mass} }
\end{figure*}

Figure \ref{fig:SFR_Mass_all} shows the spatially resolved SFR-\ma (\Ssfr - \Sm) relation considering all galaxies in the stripping sample,  using the 139727  spaxels whose emission is dominated by star formation in the galaxy disks. Here we consider together both the clumps and the diffuse emission.  The effective threshold in spatially resolved specific SFR entailed by the adopted cuts in S/N corresponds to $\rm 10^{-11.2}\, yr^{-1} \, kpc^{-2}$ \citep{Vulcani2019b} and it is also shown.
A correlation between the two quantities is immediately visible, with  spaxels with higher \Sm typically having higher values of \Ssfr. The correlation spans more than four orders of magnitude in both \Sm and  \Ssfr. Spaxels with \Sm$\rm > 10^{9} M_\odot kpc^{-2}$ form a quite thin \Ssfr - \Sm relation, which seems shifted towards lower \Ssfr values than that of the whole population. We will investigate later on who is responsible for such trend.  Overall, the scatter of the relation is $\sim0.4$ dex.  We measured this value subdividing the sample in 10 \Sm bins and computing the standard deviation of \Ssfr in each bin separately. We then took the mean value  of the standard deviations.

As comparison, Figure \ref{fig:SFR_Mass_all} also shows as density contours the relation for the 92020 disk spaxels of the galaxies belonging to the GASP control sample from \citet{Vulcani2019b}.  As discussed in that paper, the scatter of this sample is lower, being $\sim0.3$ dex. 
Overall, the datapoints of the stripping sample extend both at high and low \Sm with respect to the control sample. In addition, at any given \Sm points of stripping galaxies have systematically higher \Ssfr than  control sample galaxies. In the stripping sample, the most external contour, including 98th of the total population, systematically extends towards higher \Ssfr values than that of the control sample. In contrast, the lower edge of the same contour is very similar for the two samples. This is very similar to the effective threshold in \Ssfr we adopt and could be driven by that detection limit.  To further support the results on a statistical ground, we perform a linear regression fitting using the python module Pystan, a package for Bayesian inference. 
The parameters describing the fit, along with errors are tabulated in Tab.\ref{tab:param}, where all the fits discussed in the rest of the paper can also be found. 
Slopes are  compatible within $2\sigma$, while intercepts are different at more than $3\sigma$ level. For reference, at $\rm \log(\Sigma_\ast [M_\sun/kpc^{-2}])=8$ the $\rm \Delta(\log(\Sigma_{SFR}  [M_\sun/yr/kpc^{-2}]))$ between the two fits is $\sim 0.35$ dex.

To understand these differences, we can analyse which galaxies contribute to the different portions of the graph.

Figure \ref{fig:violin_mass} shows the distribution of \Sm as a function of the galaxy stellar mass. Data distributions are shown in terms of violin plots, which give the probability density of the data at different values, smoothed by a kernel density estimator. { Unlike bar graphs with means and error bars, violin plots contain all data points. The shape of the violin displays frequencies of values: the thicker part of the violin shape means that the values in that y-axis section of the violin have higher frequency, and the thinner part implies lower frequency. Violin plots also highlight the maximum extension of the data, and the presence of different peaks, their position and relative amplitude. The maximum width of each violin is set the same for all galaxies, for display purposes.}

The Figure clearly shows that the \Sm range depends on $M_\ast$. 
Spaxels in low mass galaxies do not reach high \Sm values:  galaxies with $M_\ast <10^{9.75} M_\odot$ have \Sm always lower than 10$\rm ^8 M_\odot kpc^{-2}$.
In contrast in massive galaxies \Sm can reach and even exceed 10$\rm ^9 M_\odot kpc^{-2}$. { At $M_\ast >10^{10.4} M_\odot$, stripping and control sample galaxies have a different behaviour: there are no control sample galaxies with \Sm $>$10$\rm ^{9.25} M_\odot kpc^{-2}$, while there are 6 stripping galaxies (15\% of the sample) in the same \Sm regime. This is only partially due to the different mass distributions of the two samples (Fig.\ref{fig:mass}): limiting the comparison to \ma$<10^{11}$ \ms we still have 5 stripping galaxies with \Sm $>$10$\rm ^{9.25} M_\odot kpc^{-2}$.} 

\begin{figure*}
\centering
\includegraphics[scale=0.35]{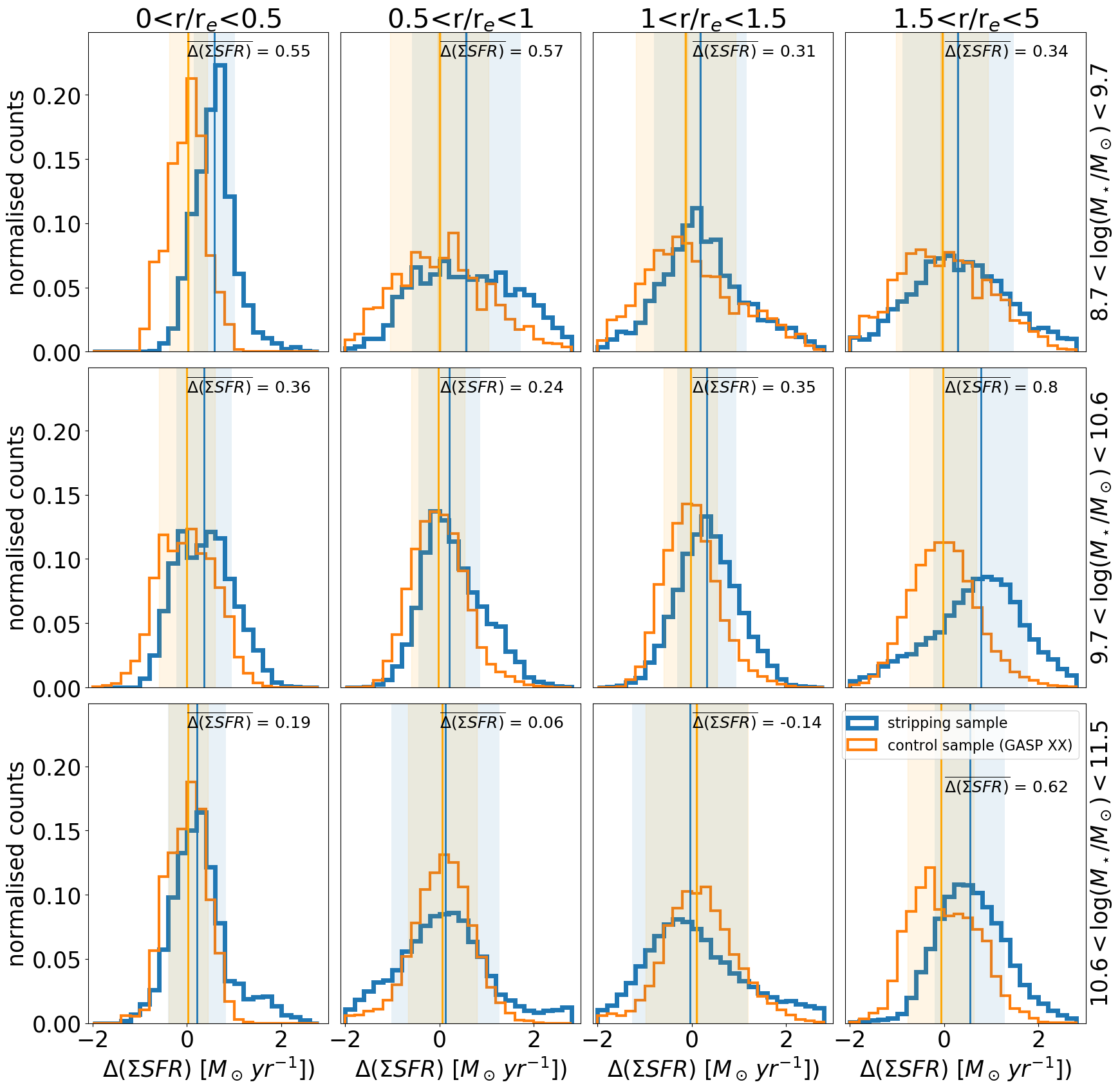}
\caption{Normalised distribution of the spaxel-by-spaxel difference between the measured \Ssfr and the \Ssfr expected from the control sample fit given the \Sm. Galaxies are subdivided into three stellar mass bins and four galactocentric distance bins, as indicated in the labels, for a total of 12 independent bins. Blue lines represent the stripping sample, orange lines the control sample. Vertical lines and shaded areas represent median values along with $1\sigma$ errors. { The difference between median values is also reported in each panel.} { Overall, the stripping sample distribution is shifted in most of the cases} towards higher $\Delta$(\Ssfr) values,  meaning that \Ssfr is enhanced with respect to the control sample, both fixing the stellar mass and the galactocentric distance. \label{fig:delta}
}
\end{figure*}

It is very important to compare samples at given stellar mass. Figure \ref{fig:delta} compares them in bins of both stellar mass and galactocentric distance, to better localise also the spatial position of the star formation enhancement. Three stellar mass bins (defined by these boundaries: $\log($\ms [$M_\odot$]) = [8.7, 9.7, 10.6, 11.5]) and four galactocentric distance bins (defined by these boundaries: r/r$_e$ = [0, 0.5, 1, 1.5, 5]) are considered, for a total of 12 independent bins.\footnote{Note that results do not change if we adopt as upper limit for the mass bins $\log($\ma [$M_\odot$]) = 11, which is the maximum mass of control sample galaxies.} In each bin, we compute  the \Ssfr-\Sm relation of the control sample galaxies in that bin and then we measure for both samples the difference between the measured \Ssfr and the \Ssfr expected from the control sample fit, given the measured \Sm. Figure \ref{fig:delta} shows the distribution of such differences. Distributions of the two samples are clearly different. In most of the cases, the stripping sample distribution is shifted towards higher values. { In each stellar mass and galactocentric distance bins,} the K-S test is able to state with very high confidence level ($p-value<<$0.01) that distributions are always drawn from different parent samples. Also median values are different in most of the cases (except for intermediate distances in the highest mass bins), when errors on medians (=1.235$\times \sigma/\sqrt{N}$ with $\sigma$ standard deviation of the distribution and $N$ number of points) are considered. These errors though are very tiny, given the high number of data points in each distribution. Figure \ref{fig:delta} shows instead the standard deviation of the distributions, which are indeed quite  broad. { Some differences in the median values between stripping and control galaxies are evident: in the galaxy central regions  (r/r$_e<$1) such difference is larger (= stripping sample has enhanced \Ssfr with respect to the control sample) among the least massive galaxies and decreases with increasing stellar mass. In the external regions, such difference does not hold anymore. Fixing the  stellar mass bin, very central  (r/r$_e<$0.5) and external regions (r/r$_e>$1.5) have a larger enhancement than intermediate regions. 
We note, however, that in the most massive bins there are only three control sample galaxies, so comparisons might not be meaningful.}{  This Figure overall suggests that in stripping galaxies the \Ssfr is enhanced with respect of the control sample, both fixing the stellar mass and the galactocentric distance. }

\begin{figure*}
\centering
\includegraphics[scale=0.52]{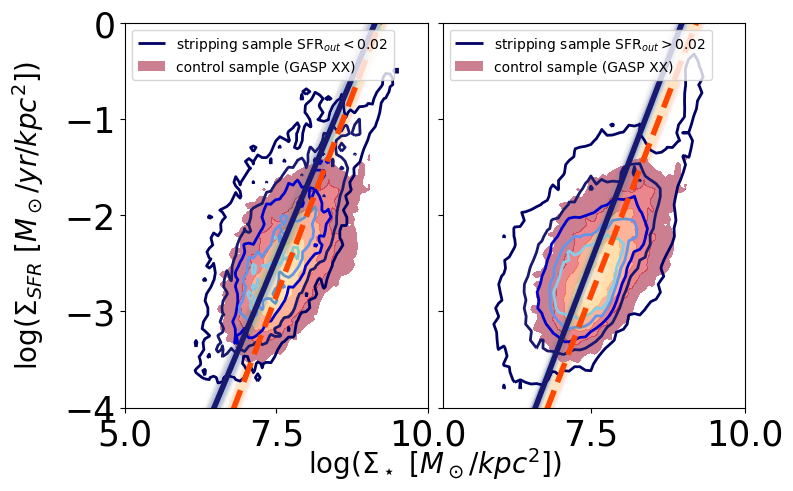}
\caption{Contour plot of the disk \Ssfr - \Sm relation for galaxies with $SFR_{out}<0.02$ (left) and with  $SFR_{out}>0.02$ (right). Colors and symbols are as in Fig.\ref{fig:SFR_Mass_all}.  Both galaxies with  low and high SFR in the tails are characterized by \Ssfr-\Sm relations shifted high compared to that of the control sample galaxies. \label{fig:SFR_Mass_SFR}}
\end{figure*}

\begin{figure*}
\centering
\includegraphics[scale=0.53]{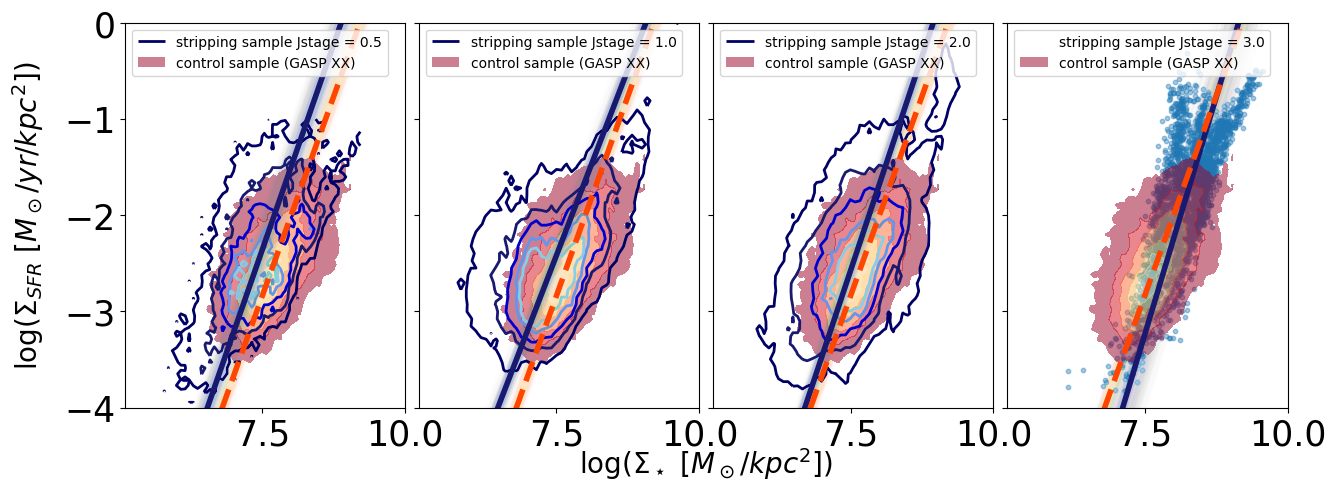}
\caption{Contour plot of the disk \Ssfr - \Sm relation for galaxies of different Jstage, as indicated in the labels.  Colors and symbols are as in Fig.\ref{fig:SFR_Mass_all}. For galaxies with Jstage=3 (last panel) no meaningful contour were possible, so all the datapoints are shown instead. The  \Ssfr enhancement is present in stripping galaxies showing any degree of stripping. Jstage=3 galaxies (truncated disks) are instead  characterised by extremely narrow relations. In their very central regions the correlation between \Ssfr and \Sm is very tight. \label{fig:SFR_Mass_Jstage} }
\end{figure*}

In addition to stellar mass, there are also other galaxy characteristics that can influence the \Ssfr-\Sm relation. As discussed in \citet{Jaffe2018, Vulcani2018_L}, and B. M. Poggianti et al. (in prep.), galaxies in the stripping sample can be categorised based on the stage of the stripping (mild, moderate and extreme, and truncated disks, see examples in Fig. 2 of \citealt{Jaffe2018}). Gullieuszik et al. (2020) also studied galaxies as a function of the amount of the SFR in the tails. We can therefore inspect the \Ssfr-\Sm relation of galaxies of these different categories, to determine whether the offset is determined by one of these groups.  Figure \ref{fig:SFR_Mass_SFR} focuses on stripping galaxies with total SFR in the tails $<0.02$ \ms yr$^{-1}$ (left) and $>0.02$ \ms yr$^{-1}$ (right), separately. Six galaxies have low level of SFR in the tails (JW29, JO138, JO23, JO197, JW108, JO10), while all the rest have $SFR_{out}>0.02$ \ms yr$^{-1}$.  Comparing these samples to the control sample, it appears evident that both galaxies with  low and high SFR in the tails are characterized by \Ssfr-\Sm relations shifted high compared to that of the control sample galaxies. Similar conclusions are reached if instead of a cut in absolute value of  SFR$_{out}$ we adopt a cut in SFR$_{out}$/SFR$_{tot}$=20\%. Linear regression fits are statistically different when comparing {
both the sample with low  and high SFR in the tails and the control sample (Tab. \ref{tab:param})}. 
Similarly, Figure \ref{fig:SFR_Mass_Jstage} shows that the  \Ssfr enhancement is present in stripping galaxies showing any degree of stripping. Galaxies with both mild, moderate and extreme stripping (Jstage = 0.5, 1, 2) show a shift of the contours towards higher \Ssfr values at any given mass, compared to the control sample contours. Galaxies with Jstage =0.5 seem to extend 
less toward high \Ssfr  at high \Sm values than the other stripping galaxies, but this might be due to fact that there are no very massive galaxies among Jstage =0.5 galaxies.
The slopes of the fits are always statistically different from that of the control sample. Galaxies with Jstage =2 seem to be the main responsible for the strip at high \Ssfr and \Sm values. 
In contrast, Jstage=3 - which are the truncated disks - show a quite different behaviour. These galaxies are characterised by extremely narrow relations. In their very central regions the correlation between \Ssfr and \Sm is very tight. In this case the fitting  parameters { describing the data of Jstage=3 galaxies} agree within the uncertainties { with those of the control sample}, but the different data distribution is outstanding. 

These results show that most of the differences between the stripping and control samples are not simply due to galaxies with very long tails and with high level of SFR in the tails, that is galaxies at the peak of the stripping, but to galaxies in all the stripping stages.
\begin{figure}
\centering
\includegraphics[scale=0.3]{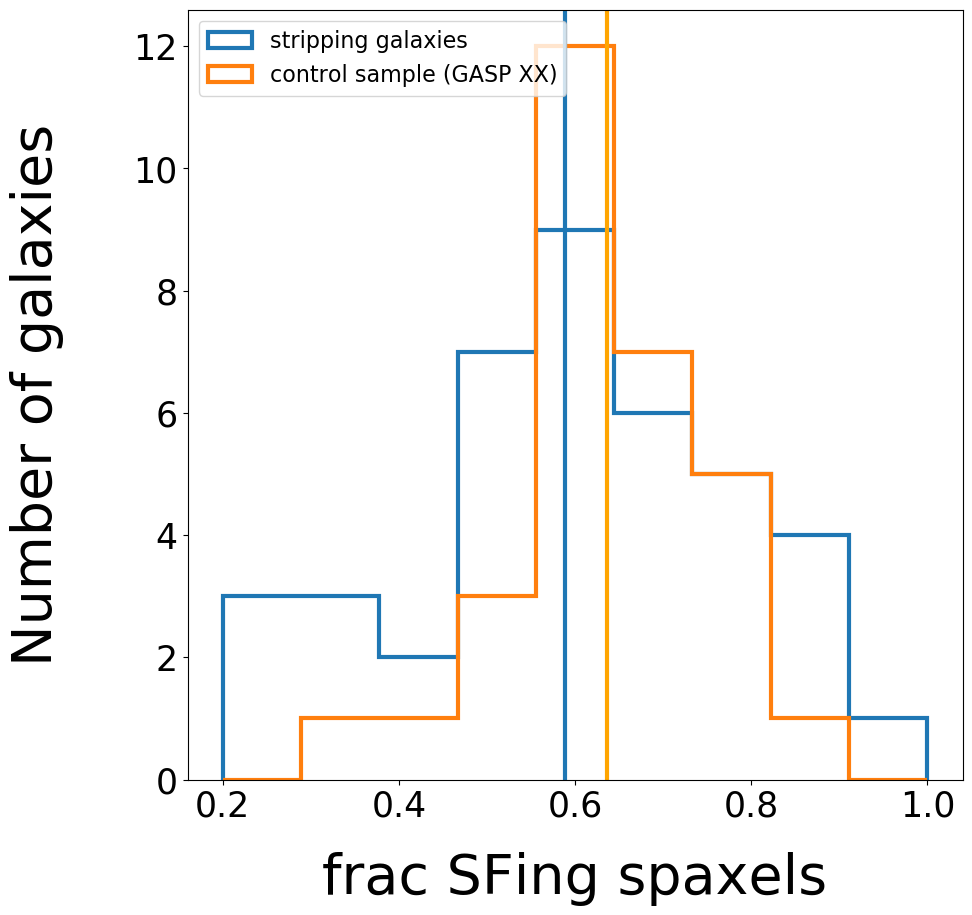}
\caption{Distribution of the fraction of star forming spaxels in galaxy disks in the stripping (blue) and control (orange) sample. Overall, the fraction of star forming spaxels within the galaxy main body is smaller in galaxies in the stripping sample.  \label{fig:frac}}
\end{figure}
This result, along with the global enhancement observed in \citet{Vulcani2018_L}, is even  more significant if we consider that in stripping galaxies the portion of the  disks which is actually star forming is much smaller than in the control sample galaxies. Fig.\ref{fig:frac} shows the distribution of the portion of galaxy disk powered by either star formation or LINER/AGN, assuming some star formation is present also in the LINER/AGN dominated regions,
for the two samples and highlights how stripping sample galaxies have overall a lower fraction
of star forming spaxels than the control sample galaxies. Excluding the spaxels powered mainly by AGN 
would only 
increase the differences.

\begin{figure*}
\centering
\includegraphics[scale=0.35]{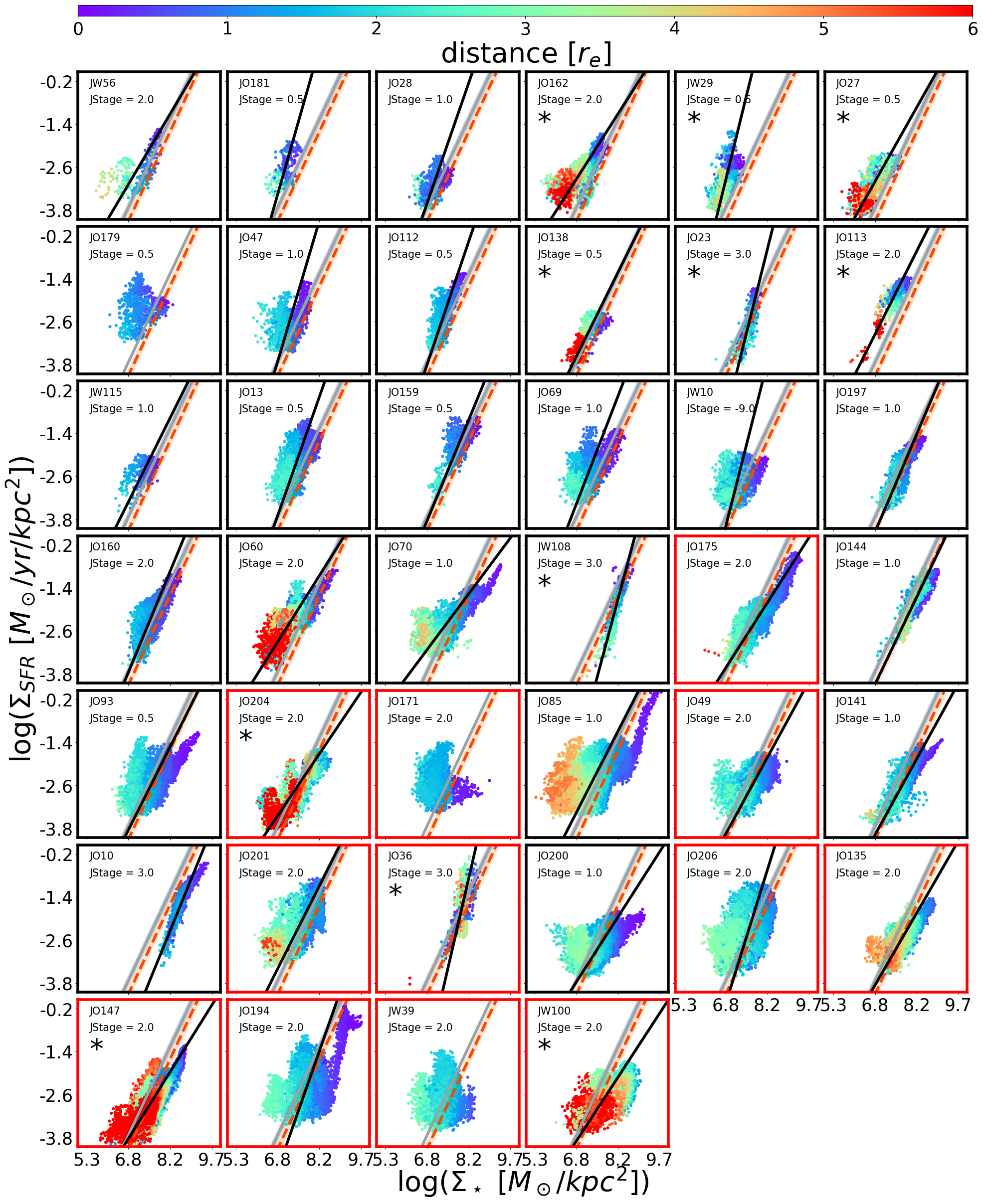}
\caption{\Ssfr - \Sm relation for all galaxies in the sample, sorted by increasing total stellar mass and colour coded by the galactocentric distance of each spaxel, in units of $r_e$. { as indicated in the color bar on the top}.  Galaxies surrounded by a red square host an AGN in their center. Galaxies labeled with an asterisk have $i>70^{\circ}$. The grey line represents the fit to the whole sample; the orange dashed line is the fit of the control sample, from \citet{Vulcani2019b}; the black line is the fit of the plotted galaxy, whose values are reported in Tab.\ref{tab:param}. Transparent lines show samples from the posterior, indicating the scatter in the fit. A reliable  fit could not be retrieved for JO179, JO171, and JW39, therefore for those galaxies there is no black line. A very large galaxy-by-galaxy variation emerges. 
\label{fig:sfr_mass_dist_mass} 
}
\end{figure*}

\subsubsection{The galaxy-by-galaxy \Ssfr- \Sm relation}

While taking into account the properties of the galaxies together allows us to study the general trends and analyse the galaxy population as a whole, it does not allow us to understand  if all galaxies follow similar relations or if each galaxy is characterized by a different slope, intercept, and scatter. 
Figure \ref{fig:sfr_mass_dist_mass} presents the \Ssfr-\Sm relation for each galaxy separately, distinguishing among spaxels at different galactocentric distances. Galaxies with high inclination ($i>70^{\circ}$) are indicated with an asterisk. Galaxies hosting an AGN are surrounded by a red square.  It appears evident that, even though overall in most cases a correlation does exist,  each object spans a distinct locus on the \Ssfr-\Sm plane: some galaxies show quite elongated sequences, some others are characterised by a cloud rather than a sequence. This is similar to what we  found in \citet{Vulcani2019b} for the control sample. Overall trends with distance are detected, with spaxels in the cores having higher values of \Sm and \Ssfr. Few cases deviates from such trend (e.g. JO36, JO27, JW29, JO147). These are most likely due to the high inclination of the galaxies that mixes spaxels at different distances and entails high levels of dust extinction.

Some galaxies have all spaxels above the total fit of the relation (e.g. JO113), while most of them have spaxels both above and below the line. Typically, especially for massive galaxies, galaxy cores are always below the fit. As seen in Fig.\ref{fig:SFR_Mass_all}, spaxels with \Sm$\rm > 10^{9} M_\odot kpc^{-2}$ have typically very thin \Ssfr-\Sm relations, probably indicating an homogeneity of the star forming properties in the galaxy cores. 
In \citet{Vulcani2019b} we showed that masking the spaxels most likely located in the galaxy  bulge -  whose size has been obtained applying a fitting on the I-band images (see A. Franchetto et al. in prep.)-  did not affect the results, showing how the suppression { of the \Ssfr} extends beyond the galaxy bulge.
As also highlighted in Fig.\ref{fig:SFR_Mass_Jstage}, Jstage=3 (JO10, JO23, JO36, JW108) follow very thin relations. All spaxels of JO10 are  well below the fit of the relation, the other 

truncated disks
cross the relation, even though they all show a quite suppressed \Ssfr given their \Sm. Note that they were not outliers in the global SFR-mass relation \citep{Vulcani2018_L}.

The presence of AGN seems not to influence the trends, but it is important to note that almost all massive galaxies in the sample host an AGN, therefore it is not possible to disentangle the two effects. 

Fitting the relation to each galaxy separately (Table \ref{tab:param}), when the fit is meaningful,\footnote{A reliable  fit could not be retrieved for JO179, JO171, and JW39.} the slope of the relation is generally different than  that of the total fit, highlighting the large galaxy-by-galaxy variation.  

\begin{figure*}
\centering
\includegraphics[scale=0.35]{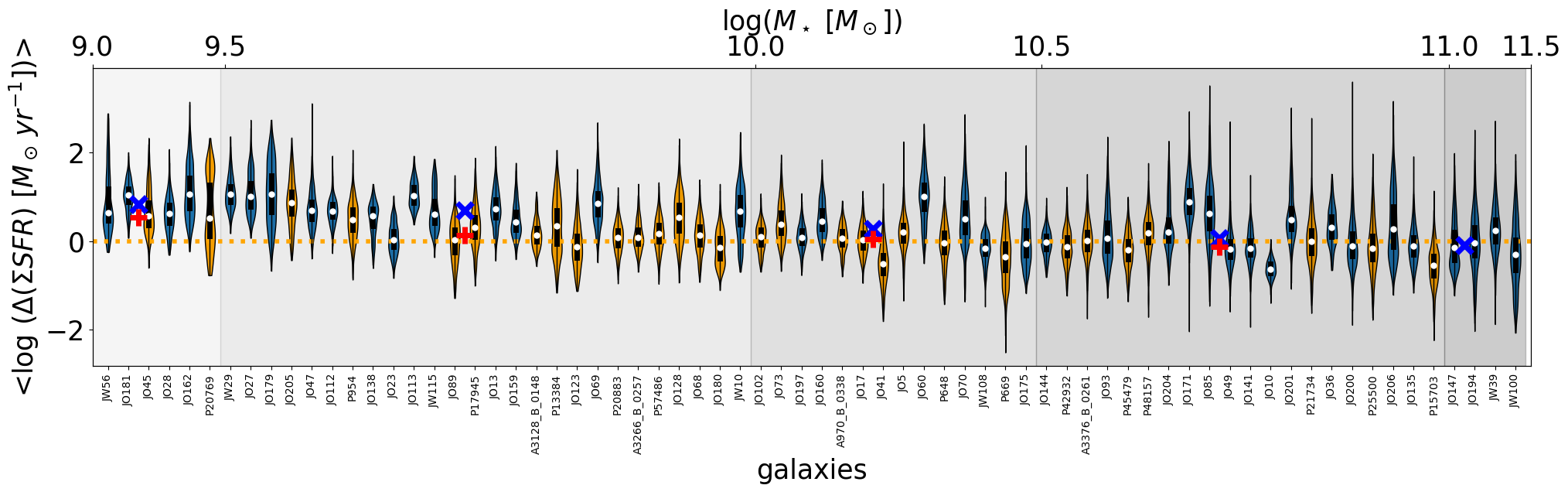}
\includegraphics[scale=0.35]{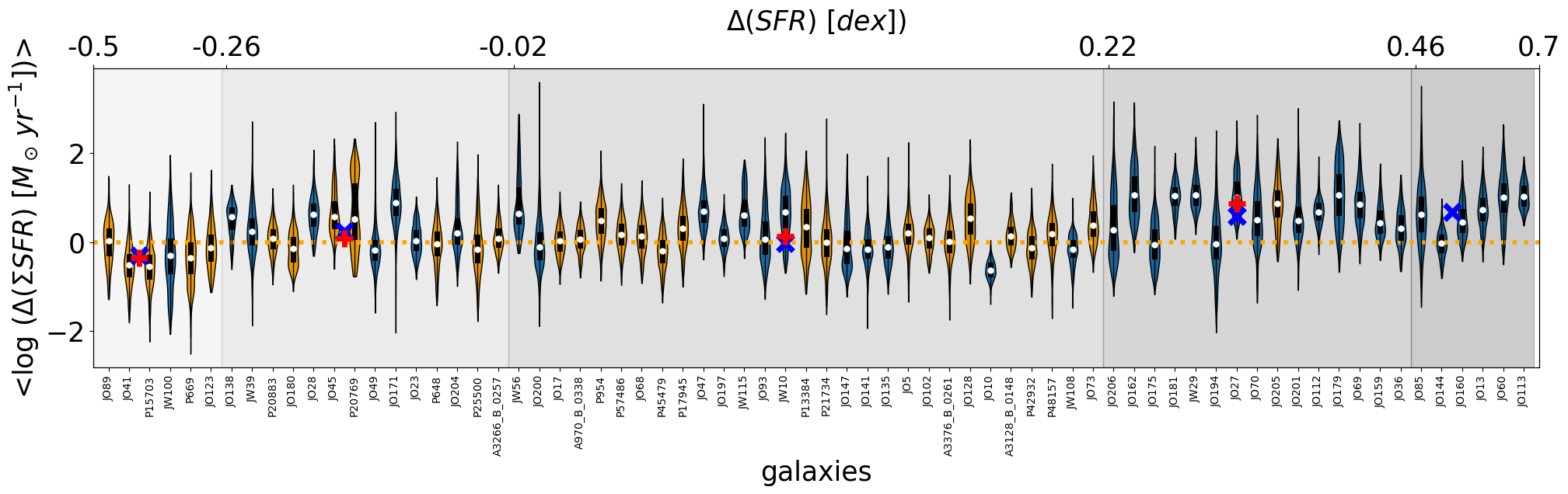}
\caption{Violin plots of the distribution of the spaxel-by-spaxel difference between the measured \Ssfr and the  \Ssfr expected from the control sample fit giving the \Sm. Galaxies are sorted by increasing stellar mass (top) and increasing $\Delta$(SFR)(bottom, from \citealt{Vulcani2018_L}), for both the control sample galaxies (orange) and the stripping galaxies (blue). Blue/red crosses represent median values in bins of stellar mass (top) and $\Delta$(SFR) (bottom) for the stripping/control sample, identified by the shaded grey areas. Dashed orange horizontal line shows the zero value, which corresponds to no offset. 
Considering all galaxies together, the median $\Delta$(\Ssfr) increases with increasing $\Delta$(SFR) and decreases with increasing stellar mass. 
At all masses stripping galaxies  have a systematically higher median  $\Delta$(\Ssfr) than their control sample counterparts. Nonetheless, the difference decreases with increasing stellar mass. In contrast, no strong differences are found between the median values of the two samples, at any $\Delta$(SFR). 
\label{fig:violin}  }
\end{figure*}

To better relate the \Ssfr and \Sm distribution of each galaxy to its global properties, we compute again the difference between the measured \Ssfr and the \Ssfr expected from the total control sample fit, given the measured \Sm. We show the distribution of the differences in  Fig. \ref{fig:violin}, using the violin plots, and we sort galaxies for increasing total stellar mass (top) and for increasing $\Delta$(SFR) (bottom). Following \citet{Vulcani2018_L},  $\Delta$(SFR) { is the difference between the SFR of each galaxy and the value derived from the control sample fit given the galaxy mass}. 
These violin plots show also the median and the interquartile ranges.
Overall, considering both samples together, the median $\Delta$(\Ssfr) increases with increasing $\Delta$(SFR) (Pearson’s correlation coefficient= 0.52,
2-tailed p-value = 3.0$\cdot10^{-6}$) and decreases with increasing stellar mass (Pearson’s correlation coefficient= -0.62,
2-tailed p-value = 1.6$\cdot10^{-8}$). Stripping galaxies not only populate the most massive end of the mass distribution (see Fig. \ref{fig:mass}), but also the highest end of the $\Delta$(SFR) distribution. The maximum value of  $\Delta$(SFR) in the control sample is 0.35 dex, in the stripping sample 0.7 dex.
Binning galaxies according to their stellar mass, we find that at all masses galaxies in the stripping sample have a systematically higher median  $\Delta$(\Ssfr) than their control sample counterparts. Nonetheless, the difference decreases with increasing stellar mass. 
In contrast, binning  galaxies according to their  $\Delta$(SFR), no strong differences are found between the median values of the two samples, at any $\Delta$(SFR). 

These results highlight a link between local and global properties of the galaxies. $\Delta$\Ssfr is influenced by stellar mass and most likely has an effect on the $\Delta$(SFR) measured on global scales.

\subsection{ \Ssfr and \Sm  properties of the \Ha clumps outside the galaxy disks in stripped galaxies}\label{sec:results_clumps}

\begin{figure}
\centering
\includegraphics[scale=0.3]{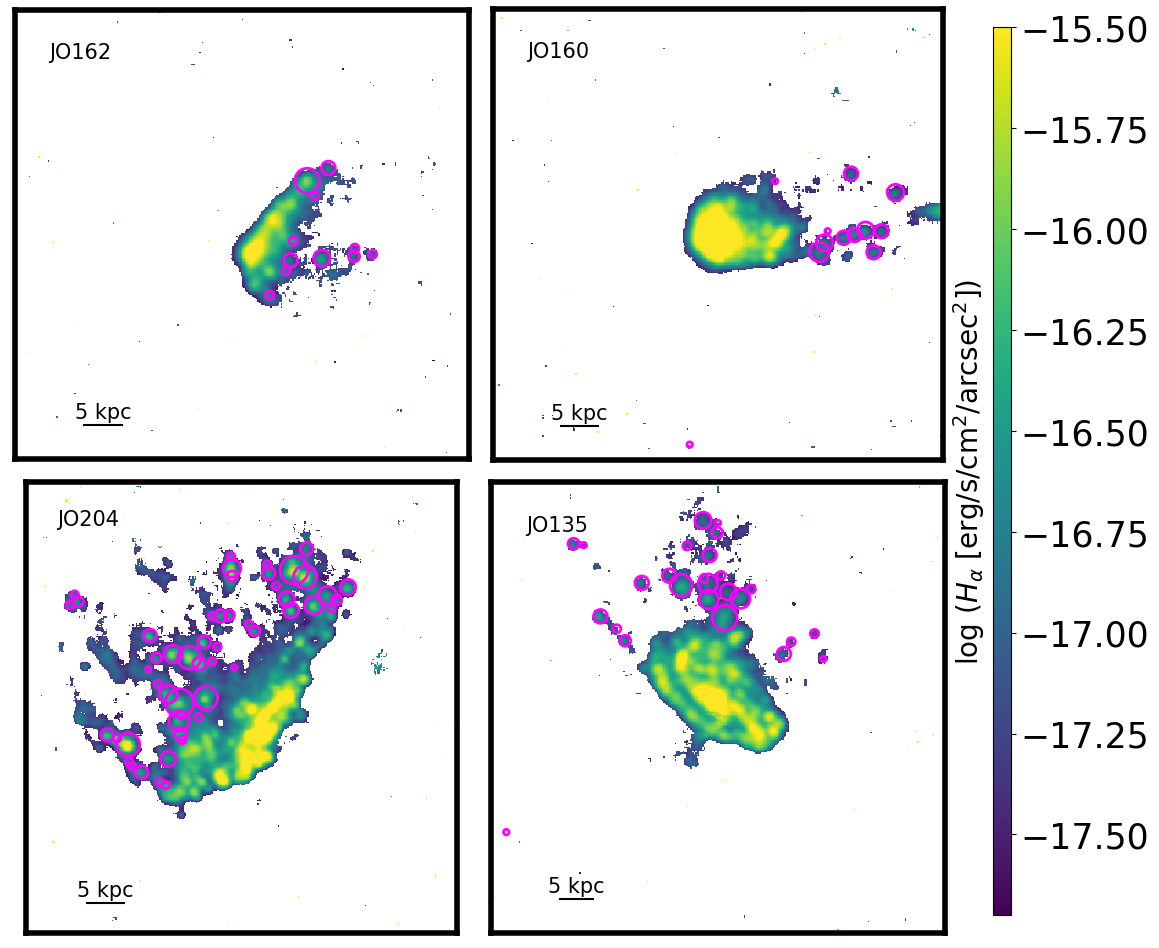}
\caption{\Ha maps for four galaxies in the stripping sample, shown as an example. Magenta circles show the \Ha clumps outside the stellar disk (i.e. in the tail).  \label{fig:Ha_map}}
\end{figure}

\begin{figure}
\centering
\includegraphics[scale=0.32]{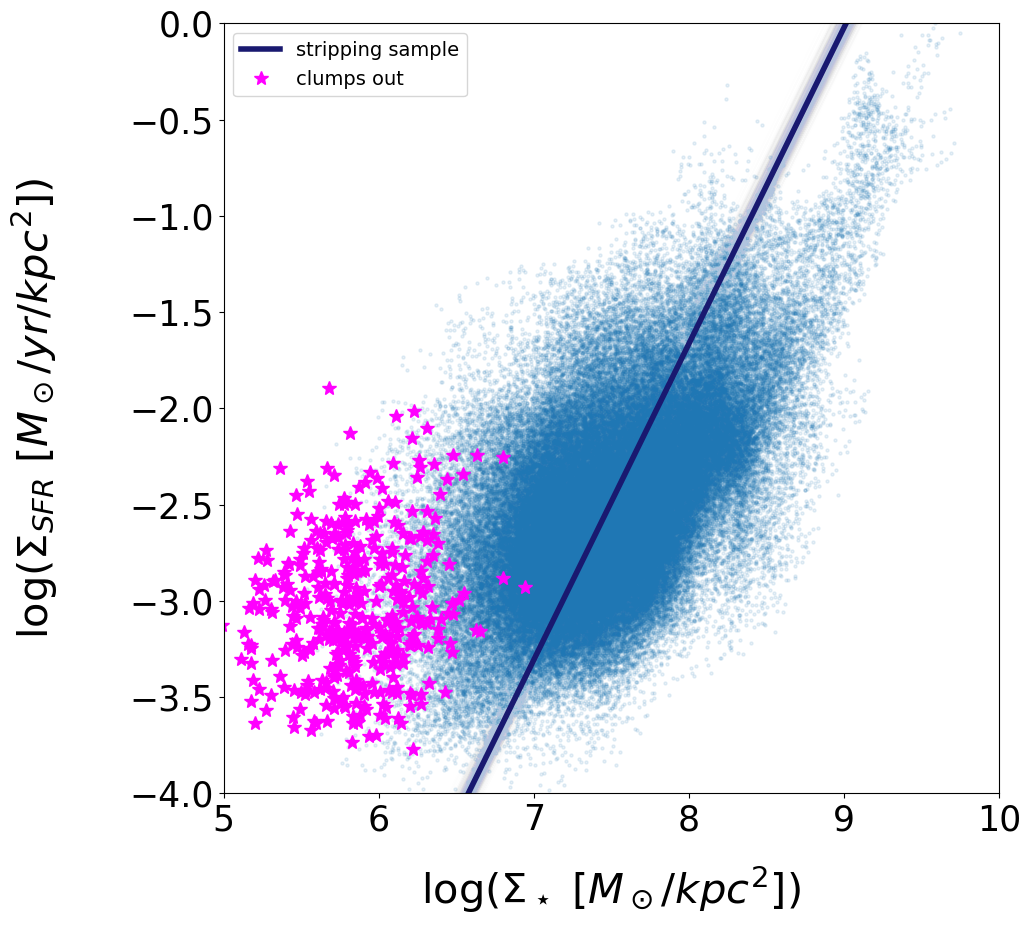}
\caption{\Ssfr - \Sm relation for all spaxels in the disks of all stripping  galaxies (blue). Superimposed  are shown the values of the \Ha clumps detected in the tails (magenta). Clumps outside the galaxy bodies do not follow a clear \Ssfr - \Sm. They have systematically higher \Ssfr values at any given  \Sm than spaxels in the galaxy disks and do not lie on the extrapolation of the disk relation.  \label{fig:blobs_sfrd}}
\end{figure}

Differently from control sample galaxies,  stripping galaxies (except for truncated disks) are characterized by the existence of material outside the galaxy disk, that is the galaxy tail. In this section we therefore focus only on the stripping sample and study the properties of the \Ha clumps detected in the tails, identified following the procedure described in Sec.\ref{sec:analysis}. 
As an example, Figure \ref{fig:Ha_map} shows the \Ha maps of four galaxies of the stripping sample.  

\citet{Poggianti2019}  characterised the properties of the clumps, considering only 16 galaxies. They found that the star forming clumps are dynamically quite cold, have a median \Ha velocity dispersion $\sigma= 27$ \kms, a median \Ha  luminosity L(\Ha)$=4\times 10^{38}\,  erg \, s^{-1}$,  a median SFR=0.003 \ms yr$^{-1}$ and \ma$=3\times10^6$ \ms. They characterised the tail clumps scaling relations (M$_{gas}$-\ms, L(\Ha)-$\sigma$, SFR-M$_{gas}$), but they did not focus specifically on the \Ssfr-\Sm relation, as we do here. 

First of all, we note that the number of star forming clumps in the tail varies from galaxy to galaxy and seems not to be strictly related to the galaxy stellar mass. Ten galaxies have no star forming clumps outside the stellar disk (JO10, JO112, JO13, JO138, JO197, JO23, JW108, JW115, JW29, JW56).

Figure \ref{fig:blobs_sfrd} shows the \Ssfr - \Sm relation for the 411 \Ha clumps in the tails, overlaid to the relation of the galaxy disks (from Fig. \ref{fig:SFR_Mass_all}). The \Sm of the clumps spans the range $10^5-10^7$ \ms kpc$^{-2}$, the \Ssfr spans the range $10^{-4}-10^{-2}$ \ms yr$^{-1}$ kpc$^{-2}$, occupying a  very different locus from that of the spaxels in the galaxy disks. A cloud rather than a well defined relation is evident. 
A Pearson’s correlation test is not able to retrieve a significant correlation (coefficient= 0.19,
2-tailed p-value = 4.0$\cdot10^{-5}$). Compared to the extrapolation of the \Ssfr - \Sm relation of the disk spaxels towards low mass surface densities, that of the clumps is shifted towards higher values at any given \Sm. We stress, however, that while the y-axis of the plots are comparable, the \Sm values of the clumps in the tails and of the spaxels in the disks have a different meaning
therefore a fair comparison is not possible. 
\Sm for the tail
clumps represent the stellar mass formed only during the ongoing star formation episode, therefore the ``true clump mass''.
In contrast, in the galaxy disks, the masses are ``projected stellar masses'' inflated by the underlying old stellar populations.

\begin{figure}
\centering
\includegraphics[scale=0.35]{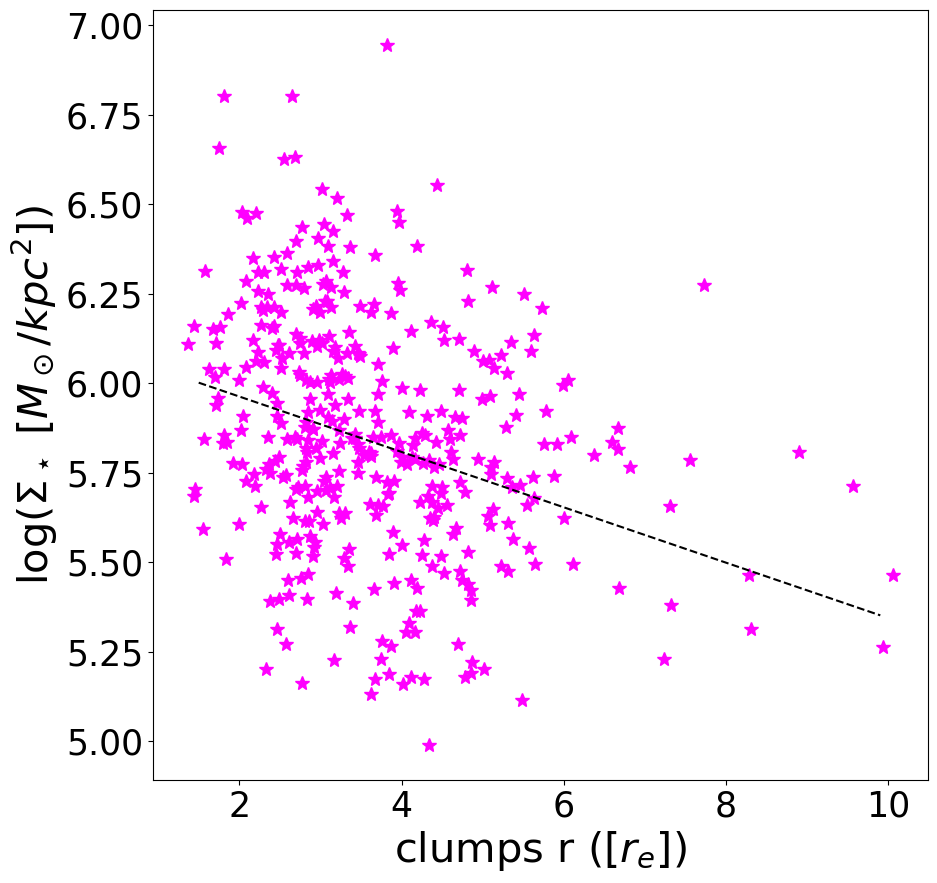}
\caption{\Sm - distance relation for all the star forming clumps in the tails of the stripping sample.  Distance is in unit of $r_e$. The black line shows the linear fit. Clumps farther away from the galaxy disks { have systematically lower \Sm values}. \label{fig:blobs_dist_m}}
\end{figure}

Overall, clumps can be found as far as 80 kpc from the galaxy center, with a median value of 25 kpc. Note that given that the clumps are extraplanar, when measuring their distance we simply compute the euclidean distance from the galaxy center, without considering the inclination to correct for projection effects. Clumps farther away from the galaxy disks are systematically less dense in mass: Fig.\ref{fig:blobs_dist_m}  shows an anticorrelation between the clump distance and  \Sm. This trend is  supported by the Pearson's correlation test  (correlation coefficient= -0.3,
2-tailed p-value = 1.0$\cdot10^{-9}$). In contrast, there seems not to be a clear correlation with \Ssfr (plot not shown).

\begin{figure}
\centering
\includegraphics[scale=0.35]{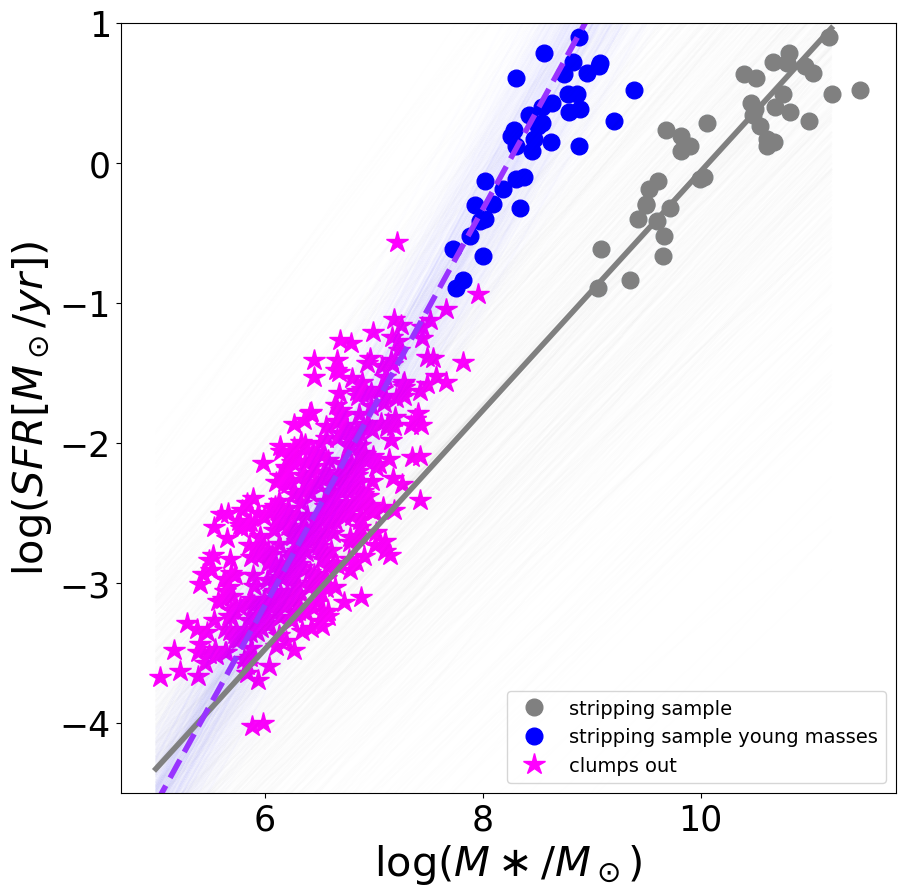}
\caption{Global SFR-mass relation for the stripping sample (grey) and the \Ha clumps found outside the disk (magenta). Blue points represent the SFR-mass relation of the galaxies when only the mass formed in the last $2\times10^8$ yr is considered (see text for details). They grey line is the fit to the global relation, the purple line is the fit representing both the magenta and blue points. Transparent lines show samples from the posterior, indicating the scatter in the fit. The relation for the clumps is shifted towards lower mass and SFR values and is much steeper when we compare the clumps and the global values for the stripping sample; it has instead the same slope when we consider only the amount of stellar mass produced only in the most recent epochs for the galaxies,  suggesting that the local mode of star formation is very similar in the galaxy disks and tails.\label{fig:blobs_sfr}} \end{figure}

As for the clumps we also have integrated values, we can compare their SFR-\ma relation to that of stripping galaxies, computed taking into account only the SFRs and masses measured within the galaxy disks. Fig. \ref{fig:blobs_sfr} shows that the relation for the clumps is not only shifted towards lower mass and SFR values, but it is also much steeper. This result might suggest that the clumps  are not simply a smaller scale of the galaxies.  However, in this comparison, stellar masses are not computed in exactly the same way. To overcome this issue, for the galaxy integrated values, we can compute the stellar mass in the same way we did for the clumps, i.e. excluding the contribution of the stellar populations older than $2\times10^8$ yr. 
In this way we can inspect the amount of stellar mass produced only in the most recent epochs. This is shown in Fig.\ref{fig:blobs_sfr} when we plot using the blue symbols only the mass formed in the last $2\times10^8$ yr: the global relation is now simply an extension of that traced by the clumps, suggesting that the local mode of star formation is very similar in the galaxy disks and tails.     

\section{discussion and conclusions} \label{sec:disc}

The analysis of the spatially resolved Star Formation Rate- Mass relation can help the understanding of how galaxies assemble at different spatial scales. Comparing the relation of galaxies located in different environments can also shed light on the role of environmental processes in enhancing on suppressing the star formation and eventually in the galaxy quenching. 

In this paper we have investigated the \Ssfr-\Sm of 40 local cluster galaxies selected for showing signs of the effects of ram pressure stripping. We have also contrasted the results with those obtained inspecting a sample of 30 undisturbed galaxies, presented in \citet{Vulcani2019b}. The 70 galaxies are drawn from the GAs Stripping Phenomena in galaxies (GASP) sample \citep{Poggianti2017}, and data have been analysed in a homogeneous way, therefore no systematic affects the  results.  

In \citet{Vulcani2018_L} we compared the integrated properties of these two samples, and found that stripping galaxies occupy the upper envelope of the undisturbed sample SFR-\ma relation, showing a systematic enhancement of the SFR at any given mass. The star formation enhancement occurs both in the disk and in the tails. In this paper we aimed at further investigating and spatially localising such SFR enhancement. 

The first result of the paper is presented in
 Figure \ref{fig:SFR_Mass_all}, which showed that even on $~\sim1$kpc scales, stripping galaxies present a systematic enhancement of \Ssfr  at any given \Sm compared to their undisturbed counterparts. This excess is as large as $\sim 0.35$ dex at \Sm=$\rm 10^{8}M_\odot \,kpc^{-2}$. This result is only partially driven by the different mass distribution between the two samples.  The excess is overall independent on the degree of stripping (except for the truncated disks) and of the amount of star formation in the tails, but is larger for less massive galaxies and decreases with increasing mass.

Interestingly, analyzing the ALMA data of a subset of the GASP galaxies, \citet{Moretti2020} have shown that galaxies undergoing ram pressure stripping have a much larger H$_2$ reservoir - which is the fuel for star formation - than normal galaxies, suggesting that this physical process  causes the conversion of large amounts of HI into the molecular phase in the disk. This result can also explain the excess in SFR that we observe.

The presence of AGNs seems not to affect the results. In addition, as the greatest differences between stripping and control sample galaxies is observed at low masses, where  galaxies in our sample do not host AGNs, results can not even be driven by a contamination of AGN spaxels not correctly identified by the BPT diagram. 

In a simple  model of a galaxy, gas is in pressure equilibrium that is set by the gravitational potential.  Thus in order for ram pressure to have any effect on this gas, it must be larger than the disk gas pressure.  This is set by the restoring force in galaxies, and often used for determining whether gas can be removed from a disk (\citet{Jaffe2018, Gullieuszik2020}.  By the same argument, gas cannot be compressed unless ram pressure is stronger than the gravitationally-set pressure.  Therefore, one would expect that because compression is relatively stronger than the gas pressure in the outskirts of galaxies, we should see a SFR enhancement in the outskirts of ram pressure stripped galaxies.  Indeed, this is seen in the simulations of \cite{Roediger2014}.

However, Fig. \ref{fig:delta}  suggests that the boost in the SFR surface density happens both in the inner and outer regions with respect to the control sample, and across a range of galaxy masses.  Therefore a simple compression argument is not as easily applied.  We argue that there are two probable causes for this.  First, as has been found in simulations \citep[e.g.,][]{TonnesenBryan2012}, dense gas that is not stripped from the outskirts of galaxies can lose angular momentum via shear from the ICM, and spiral towards the center.  This dense gas may then undergo star formation near the galaxy center, increasing the local star formation rate.  However, we may then predict that the star formation in the outskirts would decrease as dense gas migrated inward.  The second cause takes into account the varying temperature in the ISM.  Any compression wave from ram pressure may drive shocks in cold clouds (with lower sound speeds), inducing star formation.  As this does not require cloud migration it may be the more appealing picture for how star formation surface density can increase at all galactic radii.

Overall we could not detect a clear dependence on the \Ssfr-\Sm relation of the stripping stage, indicating that the \Ssfr enhancement appears as soon as the stripping begins and is maintained throughout the different stripping phases. 
Only truncated disks, representative of the final stage of stripping galaxies show a different behaviour. The fit of the relation is very similar to that of the control sample, but it is much narrower. { \cite{Fritz2017} have suggested that the existence of truncated disks points to an outside-in quenching scenario \citep[see also][]{Boselli2016}}, with the galaxy cores being the last portions of galaxies still able to produce stars. Our results though suggest that these cores are still undisturbed, and produce new stars at the same rate as undisturbed galaxies, {
similarly to what found by \cite{KoopmannKenney2004a, KoopmannKenney2004b} for a sample of galaxies in the Virgo cluster.} 

The analysis presented in this paper also highlights the existence of a large galaxy-by-galaxy variation  (Fig.\ref{fig:sfr_mass_dist_mass}), similar to that found in \citet{Vulcani2019b} for the galaxies of the control sample. In many cases, especially for the most massive galaxies, we found that galaxy cores are always below the fit, suggesting that these regions are  deprived of star formation and therefore supporting an inside-out quenching scenario according to which the suppression of the star formation occurs in the galaxy cores first and then extends to the outskirts. This behaviour is not due to the presence of a bulge \citep{Vulcani2019b} nor to the presence of an AGN. Indeed we consider only the spaxels powered by star formation.
 
A point that we did not explore here is the location of the enhancement with respect to the galaxy disk and the motion of the galaxy. In the literature, this has been done for NGC 2276, where the observed enhancement in SFR on one side of the galaxy has been explained in terms of a combination of both tidal forces and ram pressure \citep{Gruendl1993, Hummel1995, Rasmussen2006,  Wolter2015, Tomicic2018}. Inspired by this result, 
\cite{Troncoso2020}, using the EAGLE simulation, looked for the effects of the ICM on the spatially resolved star-formation activity in galaxies. They found that dividing each galaxy in two halves using the plane perpendicular to the velocity direction, differentiating the galaxy part approaching to the cluster center (the leading half), and the opposite one (the trailing half), there is an enhancement of the SFR, SFE, and interstellar medium pressure in the leading half with respect to the trailing one. Their results suggest that RP is boosting the star formation by gas compression in the leading half, and transporting the gas to the trailing half. 
As subdividing galaxies based on the velocity cut proposed by  \cite{Troncoso2020} is not feasible observationally, the same authors suggest to use instead the plane that maximizes the SFR difference, showing that it is in most cases well aligned to the velocity vector. This analysis, certainly relevant for understanding the result of this paper, is indeed deferred to a future work (I. Gaspar et al., in prep.). 
Analyzing the spatially resolved SFR of jellyfish galaxies in their initial stripping phase, they preliminarily find a SFR enhancement on the leading side (i.e. the half galaxy closest to the cluster center) of the galaxy disk. If this preliminary finding holds up, then assuming that galaxies move toward the cluster center, this relative triggering could correspond to the RPS compression.

Finally, in the last part of the paper, we have focused on the star forming clumps detected in the tails of stripping galaxies and investigated their local and global SFR-Mass relation. These clumps can be found as far as 80 kpc from the galaxy center and we detected an anticorrelation between their distance and their \Sm, indicating that further away clumps typically { have lower \Sm values}.

Investigating their \Ssfr-\Sm relation, a cloud rather than a well defined relation is evident.
The \Sm of the clumps spans the range $10^5-10^7$ \ms kpc$^{-2}$, the \Ssfr spans the range $10^{-4}-10^{-2}$ \ms yr$^{-1}$ kpc$^{-2}$, occupying a  very different locus from that of the spaxels in the galaxy disks. Compared to the extrapolation of the \Ssfr - \Sm relation of the disk spaxels towards low mass surface densities, that of the clumps is shifted towards high values at any given \Sm. We remind the reader, though, that a fair comparison is nor straightforward, as the \Sm values of the clumps in the tails represent the stellar mass formed only during the ongoing star formation episode, therefore the ``true clump mass''.
In contrast, in the galaxy disks, the masses are ``projected stellar masses'' inflated by the underlying old stellar populations. 

Considering global values, the clumps SFR-Mass relation is much steeper than that of the galaxy disk and it is not simply the extrapolation of the galaxy relation. Nonetheless, if we exclude the contribution of the stellar populations older than few $10^8$ yr in the galaxy disk values, adopting the same approach used for the clumps, the global relation becomes an extension of that traced by the clumps, suggesting that the local mode of star formation is very similar in the galaxy disks and tails. 
 
\acknowledgments
We thank the referee for their comments that helped us to improve the manuscript. Based on observations collected at the European Organisation for Astronomical Research in the Southern Hemisphere under ESO programme 196.B-0578. This project has received funding from the European Reseach Council (ERC) under the Horizon 2020 research and innovation programme (grant agreement N. 833824).  We acknowledge financial contribution from the contract ASI-INAF n.2017-14-H.0, from the grant PRIN MIUR 2017 n.20173ML3WW\_001 (PI Cimatti) and from the INAF main-stream funding programme (PI Vulcani).  


\bibliography{references}{}



\end{document}